\documentclass{aa}
\usepackage{graphicx}
\usepackage{txfonts}
\usepackage{natbib}
\newdimen\captwidth
\newdimen\figwidth
\newcommand{\mjup}{M_\mathrm{Jup}}
\newcommand{\msun}{M_\odot}

\newcommand{\rd}{\mathrm{d}}
\newcommand{\excs}{\extracolsep{\fill}}
\newcommand{\pz}{PZ~Tel~B}
\newcommand{\fomb}{Fomalhaut~b}
\newcommand{\pzc}{PZ~Tel~c}
\begin{document}
\title{Orbital fitting of imaged planetary
  companions with high eccentricities and unbound orbits}
\subtitle{Application to Fomalhaut~b and PZ Telecopii~B\thanks{ Based
    on observations collected at the European Organisation for
    Astronomical Research in the Southern Hemisphere, Chile (Program
    ID: 085.C-0867(B) and 085.C-0277(B)).}}  
\author{H. Beust\inst{1,2} \and M. Bonnefoy \inst{1,2} 
\and A.-L. Maire\inst{3} \and D. Ehrenreich\inst{4} 
\and A.-M. Lagrange\inst{1,2} \and G. Chauvin\inst{1,2}}
\institute{Univ. Grenoble Alpes, IPAG, F-38000 Grenoble, France \and
  CNRS, IPAG, F-38000 Grenoble, France \and
INAF-Osservatorio Astronomico de Padova, Vicolo dell'Osservatorio 5, 35122 Padova, Italy\and
Observatoire de Gen\`eve, Université de Gen\`eve, 51 ch. des Maillettes, CH-1290 Sauverny, Switzerland
} 
\date{Received ....; Accepted....}
\offprints{H. Beust}
\mail{Herve.Beust@obs.ujf-grenoble.fr}
\titlerunning{MCMC for high eccentricities and unbound orbits}
\authorrunning{H. Beust et al.}
\abstract{Regular follow-up of imaged companions to main-sequence
  stars often allows to detect a projected orbital motion. 
  MCMC has become very popular in
  recent years for fitting and constraining 
  their orbits. Some of these imaged companions appear
  to move on very eccentric, possibly unbound orbits. This is in
  particular the case for the exoplanet \fomb\ and the brown dwarf companion
  \pz\ on which we focus here. }
{For such orbits, standard MCMC codes assuming
  only bound orbits may be inappropriate. Our goal is to develop a new
  MCMC implementation able to handle bound and unbound orbits
  as well in a continuous manner, and to apply it to the cases of
  \fomb\ and \pz.}
{We present here this code, based on the use of
  universal Keplerian variables and Stumpff functions. 
  We present two versions of this
  code, the second one using a different set of angular variables designed to
  avoid degeneracies arising when the projected orbital motion is
  quasi-radial, as it is the case for \pz. We also present
  additional observations of \pz}
{The code is applied to \fomb\
  and \pz. Concerning \fomb, we
  confirm previous results, but we show that on the
  sole basis of the astrometric data, open orbital solutions are also
  possible. The eccentricity distribution nevertheless still peaks
  around $\sim 0.9$ in the bound regime. We present a first successful
  orbital fit of \pz, showing in particular that while both bound and
  unbound orbital solutions are equally possible, the eccentricity
  distribution presents a sharp peak very close to $e=1$, meaning a
  quasi-parabolic orbit.}
{It was recently suggested that the presence of unseen inner
companions to imaged ones may lead orbital fitting algorithms 
to artificially give very high eccentricities.
We show that this caveat is unlikely 
to apply to \fomb. Concerning \pz, we derive a
  possible solution involving an inner $~\sim 12\mjup$ companion that
  would mimic a $e=1$ orbit despite a real eccentricity around 0.7,
  but a dynamical analysis reveals that such a system would not be
  stable. We thus conclude that our orbital fit is robust.}
\maketitle
\section{Introduction}
A growing number of substellar companions are nowadays regularly
discovered and characterised by direct imaging. These objects are
usually massive (larger than a few Jupiter masses) 
and orbit at wide separations (typically $< 20\,$AU). Some of them as
sufficiently close to their host star to allow detection of their
orbital motion. Astrometric follow up gives then access to constraints
on their orbits
\citep[e.g. HR8799][]{sou11,pue15,mai16}
\citep[$\beta\:$Pictoris][]{chau12,bon14}. The
use of Markov-Chain Monte-Carlo (MCMC) algorithms to fit orbits of
substellar or planetary companions has become common now 
\citep{ford06,kal13,nel14,pue15}. This
statistical approach is particularly well suited for directly imaged
companions, as their orbit is usually only partially and unequally
covered by observations \citep{gin14}.

Some of the fitted orbits appear surprisingly very eccentric. This is
for instance the case for \fomb\ and \pz. \fomb\ is an imaged
planetary companion \citep{kal08} orbiting the A3V star Fomalhaut 
($\alpha\;$Psa, HD 216956, HIP 113368) at $\sim 119$\,AU. The physical
nature of this object is still a matter of debate. It is commonly
thought to be a low mass planet \citep{jan12,ken11,cur12,gal13}, but it also
has been suggested that the \fomb\ image could represent starlight
reflected by a cloud of dust grains, possibly bound to a real planet
\citep{kal08}. The first attempts to fit \fomb's orbit on the basis of
the available astrometric positions \citep{kal13,beu14} reveal a very
eccentric, possibly unbound orbit ($e\ga 0.8$). Subsequent dynamical
studies on the past history of this planet and its interaction with
the dust belt imaged around the star \citep{far15} led to the
conclusion that another, yet undiscovered planet must be present in
this system to control the dynamics of the dust belt, and that \fomb\
may have been formerly trapped in a mean-motion resonance with that
planet before being scattered on its present day orbit. This however
assumes that \fomb\ is actually a bound companion. While this is
likely, unbound solutions might still be possible. According to
\citet{pea15}, planets imaged of such small orbital arcs are
compatible with bound orbital solutions as well as unbound ones due to
unknown position and velocity along the line of sight.

The case of \pz\ is even more complex. PZ~Tel (HD 174429, HIP 92680),
is a G5-K8 star \citep{spe36,mes10} member of the $24\pm3\,$Myr old
\citep[][and refs. therein]{bel15} $\beta\:$Pic moving group
\citep{zuc01,tor06}. A sub-stellar companion, termed \pz, was
independently discovered by \citet{mug10} and \citet{bil10}. Its mass
is estimated $\sim 20\,\mjup$ and $\sim 40\,\mjup$
\citep{gin14,schm14}.  It is therefore most likely a substellar
object. Attempts to fit the orbit of this companion based on
successive astrometric positions led to the conclusion that it must be
close to edge-on and highly eccentric \citep{bil10,mug12,gin14}.
The pair has been imaged regularly
since 2007. \pz\ is moving away from the central star along a quasi
straight line. Its distance to the star increased by $\sim 60\,\%$
between 2007 and 2012 \citep{gin14}. From the orbital standpoint, it
is not clear whether \pz\ is a bound companion. But, in any case, its
periastron must be small ($\la 1\,$AU), which is a major difference
with \fomb. However, spectra of \pz\ obtained by \citet{schm14}
indicate that it is a young object like the star
itself. This strongly suggests that both objects are physically bound.

MCMC Orbital fitting techniques are usually based on the assumption
that the orbit to fit is elliptic, making use of corresponding
Keplerian formulas. This can be problematic with orbits with
eccentricities close to 1. This can either prevent convergence of the
fit, or generate boundaries in the fitted orbit distributions that are
not physical, but rather generated by the limitation of the method. Of
course, on could design an independent MCMC code based on the use of
open orbit formulas. Such a code would try to fit open orbits
only. Our goal is to design a code that can handle both kinds of
orbits in a continuous manner. Applied to the cases invoked above,
this would help deriving for instance a robust estimate of a
probability to be bound. This cannot be done using standard Keplerian
variables, as the changes of formulas between bound and unbound orbits
would generate enough noise to prevent the code to converge. We
develop here a MCMC code based on the use of universal Keplerian
variables, an elegant reformulation of Keplerian orbits that holds for
bound and unbound orbits as well.

The organisation of the paper is the following : In Sect.~2, we
present new VLT/NACO observations of \pz\ that we will use together with order
data in our fit. Then In Sect.~3, we present the fundamentals of our
new code based on universal Keplerian variables. In Sects.~4 and 5, we
present its application to the \fomb\ and \pz\ cases respectively. In
Sect.~6, we present further modeling based on the suggestion by
\citet{pea14} that highly eccentric companions could be actually less
eccentric that they appear due to the presence of unseen additional
inner companions. For the \pz\ case, we find one configuration that
could indeed generate an apparent very high eccentricity, but we
present subsequent dynamical modeling showing that it is in fact
unstable. Our conclusions are then presented in Sect.~7.
\section{New observations of \pz}
\subsection{Log of the observations}
\label{subsec:NaCoobs}
\pz\ was observed with VLT/NaCo \citep{len03,rou03} on September
26, 2010 in the L' band filter (central wavelength=$3.8\,\mu$m, 
width=0.62\,$\mu$m) in pupil-stabilized mode (P.I. Ehrenreich, 
Program 085.C-0277). The mode was used to subtract the stellar halo 
using the angular differential imaging (ADI) technique \citep{mar06}, 
despite the companion could be seen into our raw images.  

\begin{table*}[ht]
\caption[]{Observing log. Details on the cameras and filters can be found onto the instrument website\footnote{http://www.eso.org/sci/facilities/paranal/instruments/naco.html}.}
\label{Tab:Obs}
\begin{tabular*}{\textwidth}{@{\excs}llllllllllll}
\hline\noalign{\smallskip}
Object & Date & Band & Density filter & Camera & DIT & NDIT & 
$\mathrm{N_\mathrm{exp}}$ & 
$\mathrm{\theta_\mathrm{start}/\theta_\mathrm{end}}$ &
$\mathrm{\langle Seeing \rangle}$ & $\mathrm{\langle\tau_{0}\rangle}$ & 
Notes \\
&&&&& (s) &&& ($^{\circ}$) & (\arcsec) & (ms) & \\
\noalign{\smallskip}\hline\noalign{\smallskip}
$\theta\:$Orionis & 24/09/2010 & L' &--& L27 & 0.3 & 60 & 7 &
$-132.922/-133.862$ & 0.88 & 0.73 & Astrometric cal.\\
PZ Tel & 26/09/2010 & L' & ND\_long & L27 & 0.2 & 150 & 8 & $4.540/7.102$ & 
1.77  & 1.20& PSF \\
PZ Tel & 26/09/2010 & L' & -- & L27 & 0.3 & 100 & 143 & $10.624/53.175$ & 1.83 &
1.03 & ADI sequence\\ 
PZ Tel & 26/09/2010 & L' & ND\_long & L27 & 0.2 & 150 & 8 & $53.348/54.936$ & 
1.37 & 1.20 & PSF\\ 
\noalign{\smallskip}\hline\noalign{\smallskip}
PZ Tel & 03/05/2011 & $\mathrm{K_{s}}$ & ND\_short & S27 & 1.0 & 100 & 12 &
$-51.868/-43.155$ & 0.50 & 5.54 & ADI sequence \\
\noalign{\smallskip}\hline\noalign{\smallskip}
PZ Tel & 07/06/2011 & L' & ND\_long & L27 & 0.2 & 150 & 8 & 
$-71.868/-70.664$ & 2.78 & 0.75 & PSF \\
PZ Tel & 07/06/2011 & L' & -- & L27 & 0.2 & 150 & 96 & $-70.317/-52.769$ & 
0.81 & 2.76 & ADI sequence  \\
PZ Tel & 07/06/2011 & L' & ND\_long & L27 & 0.2 & 150 & 7 & 
$-52.582/-50.922$ & 0.78 & 2.61 & PSF \\
\noalign{\smallskip}\hline
\end{tabular*}
\end{table*}
The observation sequences, atmospheric conditions (seeing,
coherence time $\tau_{0}$), and instrument setup are summarized in
Table \ref{Tab:Obs}.  A continuous sequence of 1200
exposures was recorded, split into 8 cubes ($N_\mathrm{exp}$) of 150
images each (NDIT). The detector 512$\times$512 pixel windowing
mode was used to allow for having short data integration times
(DIT=0.3\,s). The \textit{ND\_long} neutral density (ND) was
placed into the light path for the first and last $8\times150$ frames
of the sequence to acquire unsaturated images of the star for the
purpose of the calibration of the companion photometry and
astrometry. The star core was in the non-linearity regime for the rest
of the 143 exposures. The parallactic angle ($\theta$) over the
ND-free exposures ranges from $\theta_\mathrm{start}=10.62\degr$ to
$\theta_\mathrm{end}=53.17\degr$, corresponding to an angular
variation of 2.85 times the full-width at half maximum (FWHM)
at 350\,mas.

The system was re-observed on June 7, 2011 using the same
instrument configuration (Program 087.C-0450, P.I. Ehrenreich). This
sequence was recorded under unstable conditions. Nevertheless, the
image angular resolution was high enough to resolve the brown-dwarf
companion. The observation sequence is similar to the one of
 September 26, 2010, although the field rotation is reduced
($17.55^{\circ}$). It is summarized into Table \ref{Tab:Obs}.

To conclude, we recorded one additional epoch of
pupil-stabilized observations of the PZ Tel system in the $K_{s}$
band (central wavelength=2.18 $\mu$m, width=0.35 $\mu$m) with NaCo
(P.I. Lagrange, Program 087.C-0431). We used the neutral density
filter of the instrument (\emph{ND\_Short}) adequate to this band to avoid
saturating the star.  The field rotation was not sufficient to take
advantage of the angular differential imaging technique.
\subsection{Data Reduction}
The reduction of the L'-band  observations was carried out by 
a pipeline developed in Grenoble\citep{bon11,chau12}.
The pipeline first applied the basic cosmetic steps (bad
pixel removal, nodded frame subtraction) to the raw images. The star
was then registered into each individual frames of each cube using a
2D Moffat function adjusted onto the non-saturated portions of the
images. We applied a frame selection inside each cube based on the
maximum flux, and on the integrated flux over the PSF core. The cubes
were then concatenated to create a master cube.

Given the relative brightness of the companion and the amount of field
rotation for the 2010 observations, we chose to apply the smart-ADI
(SADI) flux-conservative algorithm to suppress the stellar halo
\citep{mar06}. The algorithm builds for each frame of our
observation sequence a model of the stellar halo from the other
sequence images for which a companion located at a distance $R$ from
the star has moved angularly (because of the field rotation) by more
than $n \times$ the FWHM (separation criterion). Only the $NN$
($NN\in R$) frames the closest in time (Depth parameter) and
respecting the separation criterion are considered. We defer the
reader to \citet{mar06} and \citet{bon11} for details. We found the
parameters maximizing the detection significance of the companion
($R=$13.6 pixels, depth=4, and 2$\times$FWHMs) throughout these
intervals: $2\leq\mathrm{Depth}\leq 12$, $\mathrm{FWHM=}1, 1.5, 2$.

Flux losses associated to the self-subtraction of the companion
during the ADI process were estimated using either five artificial
planets (AP) are injected at $\mathrm{PA}=179\degr$, $-61\degr$,
$239\degr$, $210\degr$, and $270\degr$, or negative fake planets
\citep{bon11}. These AP were built from the non-saturated exposures of
the star taken before and after the ADI observations (see
Table~\ref{Tab:Obs}).

We derive a final contrasts of $\Delta L^{\prime}=5.15\pm0.15$\,mag. The error
accounts for uncertainties on the flux losses estimates, on the
evolution of the PSF through the ADI sequence, and on the method used
to extract the companion photometry. This new L' band photometry was
accounted for the up-to-date analysis of the spectral energy
distribution of the companion \citep{mai16}.

Images of the $\theta$ Orionis C astrometric field were acquired with
an identical setup on September, 24 2010. They were properly reduced
with the \texttt{Eclipse} software \citep{dev97}. The position
of the stars on the detector were compared to their position on sky
measured by \cite{mac94} to derive the instrument orientation to
the North of $-0.36\pm0.11\degr$ and a detector plate scale of
$27.13\pm0.09$\,mas. We used these measurements together with the
position of \pz\ derived from the negative fake planet injection
\citep{bon11} to find a PA=$59.9\pm0.7\degr$ and a separation of
$374\pm5$\,mas for the companion.

The second epoch of L' band observations was reduced with the
  IPAG pipeline, but using the classical imaging (CADI)
  algorithm. The algorithm built a model of the halo valid for all
  the images of the sequence from the median of all images contained
  in the sequence. It is more appropriate than the smart-ADI
  algorithms here because of the small amount of field
  rotation. Indeed, it would not have been possible to built a model
  of the halo for each frames of the sequence while respecting a
  separation criterion of 1 FWHM for all the frames contained in our
  sequence.  The flux losses were estimated in the same way as for the
  2010 L'-band observations. We measure a contrast of $\Delta
  L^{\prime}=4.6\pm0.6$\,mag. The photometry is less accurate because
  of the unstable conditions during the observations.

The instrument orientation (TN=$1.33\pm0.05^{\circ}$) and plate-scale 
($27.38\pm0.08$ mas/pixel) were measured onto the images of the binary star 
HD 113984 \citep{vand93} observed on September 2, 2011. We  therefore derive 
a PA=$60.0\pm0.6^{\circ}$ and separation of $390.0\pm5.0$ mas for the 
companion.

We realigned each of the $K_{s}$ band images to the North and
  median-combined them to create a final image of PZ Tel AB. The
  companion is seen in the images. We removed the stellar halo at the
  position of the companion making a axial symmetry of the halo around
  a axis inclined at PA=-45, 0, or 90$^{\circ}$.  We integrated the
  flux of the star and companions into circular apertures of radii
  135 mas (5 pixels) to derive a contrast ratio of $\delta
  K_{s}=5.46\pm0.07$. The error bars account for the dispersion of
  contrast found for the different choice of duplication axis for the
  stellar halo removal. This contrast ratio is consistent within error
  bars with the one derived by \citet{mug12} with the same instrumental
  setup. We measure a PA=$60.0\pm0.6^{\circ}$ and a separation of
  $390.0\pm5.0$ mas for the system. This astrometry relies on the TN
  and plate-scale measured on March 03, 11, and reported in
  \citet{chau12}. The astrometry reported in this section assumes that
  the TN and plate-scale are stable between the observations of the
  astrometric fields and our observations of PZ Tel. This seems to be
  the case according to the measurements of \citet{chau12}.
\section{Fundamentals of MCMC for high eccentricity 
and open orbits}
\subsection{MCMC for astrometric imaged companions}
The fundamentals of the MCMC method applied to planets detected 
with radial velocities are described in \citet{ford05,ford06},
and its application to imaged companions are for
instance described in \citet{chau12}. The first requirement is to
presuppose general probability distributions (commonly called priors)
for the orbital elements. For bounds orbits, the usual orbital
elements are the semi-major axis $a$, the eccentricity $e$, the
inclination $i$, the longitude of ascending node $\Omega$, the
argument of periastron $\omega$, and the time for periastron passage
$t_p$. The priors for these elements are generally assumed uniform
between natural bounds, except for $a$ for which a logarithmic prior
($\propto\ln a$) is often assumed, and for $i$ for which assume a
prior $\propto\sin i$ is also of standard use. Combined with uniform
prior for $\Omega$, this choice ensures a uniform probability
distribution over the sphere for the direction of the orbital angular
momentum vector.

In the building process of a Markov chain, successive orbital
solutions are generated from preceding ones taking steps on the
orbital variables and selecting or rejecting the generated new orbits
using the Metropolis-Hastings algorithm (hereafter MH)
\citep{ford05}. MCMC theory tells that whenever the chain grows, it is
expected to stabilise and the final statistical distribution of orbits
within the chain samples the posterior probabilistic distribution of
orbital solutions. In practice several chains are run in parallel (we
use 10), and Gelman-Rubin criterion on crossed variances is used to
check for convergence \citep{ford06}.

An important point to note is that the variables on which steps are
taken with MH are not necessarily the orbital elements listed above
themselves, but rather combinations of them. In \citet{chau12}
($\beta\,$Pic b) and \citet{beu14} (\fomb), the work is done on
the parameter vector
\begin{eqnarray}
\vec{w_1} & = & \left(\frac{\cos(\omega+\Omega+v_0)}{P},
\frac{\sin(\omega+\Omega+v_0)}{P},\frac{e\cos(\omega+\Omega)}{\sqrt{1-e^2}},
\right.\nonumber\\
&&\frac{e\sin(\omega+\Omega)}{\sqrt{1-e^2}},
(1-e^2)^{1/4}\sin\frac{i}{2}\cos(\omega-\Omega),\nonumber\\
&&\left.(1-e^2)^{1/4}\sin\frac{i}{2}\sin(\omega-\Omega)\right),
\label{param1}
\end{eqnarray}
where $P$ is the orbital period and $v_0$ is the true anomaly at a
reference epoch (typically that of a specific data point of a time
close to periastron passage). This choice was dictated by the
following considerations~:
\begin{itemize}
\item As pointed out in \citet{chau12}, considering imaged
  companions, there is a degeneracy in orbital solutions concerning
  parameters $\Omega$ and $\omega$. Considering one potential solution
  with $(\Omega, \omega)$ values, the same solution but with
  $(\Omega+\pi, \omega+\pi)$ exactly gives the same projected orbital
  motion. The only way to lift this degeneracy is to have independent
  information about which side of the projected orbit (or of the
  associated disk) is on the foreground, or to have radial velocity
  measurements. Hence taking steps on $\Omega$ and $\omega$ in MCMC
  can generate convergence difficulties with chains oscillating
  between two symmetric families of solutions. To avoid this
  difficulty, we consider angles $\omega+\Omega$ and $\omega-\Omega$
  which are unambiguously determined. It is indeed easy to express the
  projected Keplerian model as a function of these angles only.
\item Whenever $i=0$, angles $\Omega$ and $\omega$ are undefined (and
  so angle $\omega-\Omega$), while $\Omega+\omega$ is still
  defined. Hence we take variables $\propto\sin(i/2)\cos(\omega-\Omega)$ and
  $\propto\sin(i/2)\sin(\omega-\Omega)$ to avoid a singularity
  whenever $i\rightarrow 0$.
\item The same applies to eccentricity. When $e$ vanishes,
  $\Omega+\omega$ itself is undefined. This is why we consider
  variables $\propto e\cos(\Omega+\omega)$ and $\propto
  e\sin(\Omega+\omega)$.
\item $\omega+\Omega+v_0$ is defined even when $e=i=0$. This is why we
  chose it in the remaining variables. But as much as possible, we
  avoid directly taking steps on angular variables themselves, which
  can lead to convergence problems with jumps around $2\pi$ and
  similar values. This is why we use $\cos(\omega+\Omega+v_0)$ and
  $\sin(\omega+\Omega+v_0)$ in the remaining variables.
\end{itemize}
As explained by \citet{ford06}, the assumed priors are taking into
account in MH multiplying the basic probability by the Jacobian of the
transformation from the linear vector $(\ln a, e, \sin i,\Omega,
\omega, t_p)$ to the parameters vectors. This Jacobian reads here
\begin{equation}
J_1=\frac{1}{2}\frac{e(1+e\cos v_0)}{(1-e^2)^3P^2}\qquad.
\end{equation}
\subsection{Open orbits and universal variables}
The parameter vector $\vec{w_1}$ (\ref{param1}) is well suited to fit low
eccentricity orbits. It has nevertheless proved efficiency for high
eccentricity orbits as well \citep{beu14}. \citet{ford06} also gives
alternate sets of parameters for such orbits. However, none of them
can handle open orbits. Moreover, their validity of the fit close to
the boundary $e=1$ is questionable. Our goal is to allow MCMC fitting
to account for bound or unbound orbits in a continuous manner as
well. Some of the variables in Eqs.~(\ref{param1}), such as the
orbital period $P$ and $\sqrt{1-e^2}$ are clearly inappropriate and
need to be changed. The periastron $q$ is conversely always defined
for any type of orbit. So changing $P$ to $q$ and eliminating
$\sqrt{1-e^2}$ in Eqs.~(\ref{param1}) could be a first
solution. We designed a code based on this idea, which turned out not
to be efficient. Convergence of Markov Chains could not be reached
after billions of iterations. The reason lies in the Keplerian model
assumed. To be able to compute the position and velocity of an
orbiting companion at a given time (and subsequently a $\chi^2$), one
needs to solve Kepler's equation for the eccentric anomaly $u$ as a
function the time $t$. Kepler's equation depends on the type of
orbit. For an elliptical, parabolic and hyperbolic orbit, this
equation reads
\begin{equation}  
u-e\sin u = M,\qquad\frac{u}{2}+\frac{u^3}{6}=M,\quad\mbox{and}
\quad e\sinh u-u = M 
\label{keps}
\end{equation}
respectively. In the parabolic case, this equation is called Barker's
equation, and $u=\tan(v/2)$, where $v$ is the true
anomaly. $M=n(t-t_p)$ is the mean anomaly, while $n$ is the mean
motion. This equation holds in all cases, but $n$ is defined as
$\sqrt{\mu/a^3}$ in the elliptic and hyperbolic case, and as
$\sqrt{\mu/8q^3}$ in the parabolic case, where $\mu=GM$ is the
dynamical mass ($M$ is the central mass). In the random walk process
of a Markov chain, permanent switching between these equations lead to
instabilities that prevent convergence. A good way to overcome this
difficulty is to move to the universal variable formulation
\citep{dan83,bur83,dan87}, which is a very elegant way to provide a
unique and continuous alternative to Kepler's equation valid for any
kind of orbit. We first define the energy parameter $\alpha$ as
\begin{equation}
\alpha=-2E=\frac{\mu}{q}\,(1-e)\qquad,
\end{equation}
where $E$ is the energy per unit mass. This expression is valid for
any orbit. Elliptical orbits are characterized by $\alpha>0$,
parabolic orbits by $\alpha=0$ and hyperbolic ones by $\alpha<0$. The
eccentric anomaly $u$ is then replaced by the universal variable
$s$. For non-parabolic orbits, $s$ is defined as
\begin{equation}
s=\frac{u}{\sqrt{|\alpha|}}\qquad,
\end{equation}
and as
\begin{equation}
s=\frac{u}{2qn}
\end{equation}
for parabolic orbits. It can be shown that in any case, we have
\begin{equation}
s=\frac{1-e}{q}\,\left(t-t_p\right)+\frac{eY}{\sqrt{q\mu(1+e)}}\qquad,
\end{equation}
where $Y$ is the $y$-coordinate in a local $OXY$ referential frame
($X$ axis pointing towards periastron). This shows that the definition
of $s$ is continuous irrespective of the type of orbit. Then,
\citet{bur83} show that Kepler's equation can then be rewritten in any
case as
\begin{equation}
\mu s^3c_3\left(\alpha s^2\right)+qsc_1\left(\alpha s^2\right)=t-t_p\qquad,
\label{kepu}
\end{equation}
where $t$ is the time, and the $c_k$'s are the Stumpff functions defined as
\begin{equation}
c_k(x)=\sum_{n=0}^{+\infty}\frac{(-1)^n}{(2n+k)!}\,x^n\qquad.
\label{ck}
\end{equation}
This formulation of Kepler's equation is valid for any orbit. Using 
$c_k(0)=1/k!$, we see that for $\alpha=0$ (parabolic orbit), this equation 
is equivalent to Barker's equation. Once this equation is solved 
for $s$, the heliocentric distance $r$ and the rectangular 
coordinates $X$ and $Y$ read
\begin{eqnarray}
X & = & q-\mu s^2c_2\left(\alpha s^2\right)\qquad,\\
Y & = & s\sqrt{q\mu(1+e)}\,c_1\left(\alpha s^2\right)\qquad,\\
r & = & q+e\mu s^2c_2\left(\alpha s^2\right)
\qquad,
\end{eqnarray}    
these formulas applying for any type of orbit. 
This formalism was used by several authors for specific problems,
such as \citet{aar99} for wide binaries in clusters, or \citet{cab01} to 
sole Keplerian problems with additional disturbing potentials $\propto 1/r^2$. 
The Kepler advancing routines in the popular symplectic $N$-body packages 
\textsc{Swift} \citep{ld94} and \textsc{Mercury} \citep{cham99} are also 
coded this way for high eccentricity and open orbits. Very recently, 
\citet{wis15} proposed an alternate formalism that avoids the use of 
Stumpff functions. They claim it to be more efficient. 
We did not try to apply that yet and used the Stumpff functions theory 
instead.

Based on the use of Stumpff functions, 
we designed a new MCMC code, using the following parameter vector
\begin{eqnarray}
\vec{w_2} & = & \left(\rule[-1.5ex]{0pt}{3ex}
q\cos(\omega+\Omega), q\sin(\omega+\Omega),\right.
\nonumber\\
&& \qquad\left.\sin\frac{i}{2}\cos(\omega-\Omega),
\sin\frac{i}{2}\sin(\omega-\Omega),e,s_0\right)\quad,
\label{param2}
\end{eqnarray}
where $s_0$ is the universal variable at a given reference epoch. 
The priors are assumed uniform for $\Omega$, $\omega$, $e$, and $t_p$,
logarithmic for $q$ and $\propto\sin i$ for $i$. The Jacobian of the 
transformation from  $(\ln q,e,\sin i,\Omega,\omega,t_p)$ to $\vec{w_2}$ reads
\begin{equation}
J_2=\frac{q^2}{2}\,\frac{\rd s}{\rd t}\qquad,
\end{equation}
where $\rd s/\rd t$ can be obtained as
\begin{equation}
\frac{\rd s}{\rd t}=\frac{1}{\mu s^2c_2\left(\alpha s^2\right)
+qc_0\left(\alpha s^2\right)}\qquad,
\end{equation}
to be evaluated here at $s=s_0$.
\subsection{Transformation of angles}
This new code was able to reach convergence in the \fomb
case. Convergence appeared however hard to reach in the PZ Tel
case. This is due to the structure of the data (Table~\ref{pzteldata}). The
astrometric data of \pz\ reveal indeed a quasi-linear motion
nearly aligned with the central star. \pz's orbit appears
extremely eccentric, perhaps unbound, but with a periastron much
smaller ($\la1\,$AU) than the measured projected distances. In this
context, the local reference frame $OXYZ$ may be not well
defined. Only the line of apsides $OX$ is likely to be well
constrained by the data, while the two other directions require an
other nearly arbitrary angular variable to be fixed. The very bad
constraint on that angular variable results into degeneracies in the
constraints on angles $(\Omega,\omega,i)$ that are sufficient to
prevent convergence. It thus appears necessary to isolate the badly
constrained angular variable into a specific variable. This require to
change the parameter vector $\vec{w_2}$ (Eq.~\ref{param2}).

With respect to the sky reference frame ($x$-axis pointing towards north,
$y$-axis towards east, and $z$-axis towards the Earth), the basis vectors 
($\vec{e_X},\vec{e_Y},\vec{e_Z}$) of the local $OXYZ$ reference frame read
\begin{eqnarray}
\vec{e_X}&&\left\{
\begin{array}{l}
\cos\omega\cos\Omega-\cos i\sin\omega\sin\Omega\\
\cos\omega\sin\Omega+\cos i\sin\omega\cos\Omega\\
\sin\omega\sin i
\end{array}\right.\qquad,\nonumber\\
\vec{e_Y}&&\left\{
\begin{array}{l}
-\sin\omega\cos\Omega-\cos i\cos\omega\sin\Omega\\
-\sin\omega\sin\Omega+\cos i\cos\omega\cos\Omega\\
\cos\omega\sin i
\end{array}\right.\qquad,\nonumber\\
\vec{e_Z}&&\left\{
\begin{array}{l}
\sin i\sin\Omega\\
-\sin i\cos\Omega\\
\cos i
\end{array}\right.\qquad.
\label{exeyez}
\end{eqnarray}
According to our analysis, in the case of \pz, only $\vec{e_X}$
is well constrained by the data. This results in complex combined
constraints on $\Omega$, $\omega$ and $i$. For instance, if
$\vec{e_Z}$ was well constrained by the data, then $\omega$ would be
the weakly constrained parameter, as $\vec{e_Z}$ is only function of
$\Omega$ and $i$. The idea is therefore to define new angles in a
similar way as in Eqs.~(\ref{exeyez}), but in such a way that $\vec{e_X}$
depends on two angles only. We thus introduce new angles $i'$,
$\Omega'$ and $\omega'$ designed in such a way that $\vec{e_X}$ is
defined with respect to ($i',\Omega',\omega'$) in the same manner as
$\vec{e_Z}$ is defined with respect to ($i,\Omega,\omega$). Similarly
$\vec{e_Y}$ will be defined like $\vec{e_X}$, and $\vec{e_Z}$like
$\vec{e_Y}$. We thus write now
\begin{eqnarray}
\vec{e_X}&&\left\{
\begin{array}{l}
\sin i'\sin\Omega'\\[\jot]
-\sin i'\cos\Omega'\\[\jot]
\cos i'
\end{array}\right.\qquad,\nonumber\\
\vec{e_Y}&&\left\{
\begin{array}{l}
\cos\omega'\cos\Omega'-\cos i'\sin\omega'\sin\Omega' \\
\cos\omega'\sin\Omega'+\cos i'\sin\omega'\cos\Omega' \\
\sin'\omega\sin i' 
\end{array}\right.\qquad,\nonumber\\
\vec{e_Z}&&\left\{
\begin{array}{l}
-\sin\omega'\cos\Omega'-\cos i'\cos\omega'\sin\Omega'\\
-\sin\omega'\sin\Omega'+\cos i'\cos\omega'\cos\Omega'\\
\cos\omega'\sin i'
\end{array}\right.\qquad.
\label{exeyez2}
\end{eqnarray}
The comparison between formulas (\ref{exeyez}) and (\ref{exeyez2})
gives the correspondence between the two sets of angles. Now, the line
of apsides is defined by $i'$ and $\Omega'$ only, and $\omega'$, the
badly constrained angular variable, is undefined if $q$ vanishes. It
is therefore worth modifying vector $\vec{w_2}$ according to this
transformation. However, one should not forget that vector
$\vec{w_2}$ was designed to avoid the natural degeneracy of
astrometric solution between ($\Omega, \omega$) and ($\Omega+\pi,
\omega+\pi$). It can be seen from Eqs.~(\ref{exeyez}) and
(\ref{exeyez2}) that changing ($\Omega, \omega$) to ($\Omega+\pi,
\omega+\pi$) is equivalent to changing ($i', \omega'$) to ($\pi-i',
-\omega'$) while leaving $\Omega'$ unchanged. This transformation
leaves the first two components of $\vec{e_X}$ and $\vec{e_Y}$
unchanged (this explains the degeneracy of the projected orbit), as
well as the third component of $\vec{e_Z}$, all remaining components
being changed to their opposites. The new parameter vector must remain
unchanged with this transformation as well, to avoid convergence
difficulties. We chose the following new parameter vector
\begin{eqnarray}
\vec{w_3} & = &  \left(\sin i'\cos\Omega', \sin i'\sin\Omega',\right.
\nonumber\\
&& \qquad\left.
q\sin i'\cos\omega', q\cos i'\sin\omega',e,s_0\right)\qquad.
\label{param3}
\end{eqnarray}
This new vector is invariant in the transformation $(i',
\omega')\longrightarrow(\pi-i', -\omega')$. Its first two components
define $\vec{e_X}$ unambiguously. Its third and fourth components
vanish when $q=0$, i.e., when $\omega'$ is undefined, which avoids
singularities. Now, this vector can be expressed as a function of the
original angles ($i,\Omega,\omega$) directly, so that the formal
introduction of the new angles ($i',\Omega',\omega'$) is
unnecessary. The same vector can be written
\begin{eqnarray}
\vec{w_3} & = &  \left(\rule[-2.5ex]{0pt}{5ex}
-\cos\omega\sin\Omega-\sin\omega\cos i\cos\Omega,\right.\nonumber\\
&&\qquad\cos\omega\cos\Omega-\sin\omega\cos i\sin\Omega,\nonumber\\
&&\qquad\qquad\left.q\,\cos i,q\,\frac{\cos\omega\sin\omega\sin^2i}
{\sqrt{1-\sin^2i\sin^2\omega}},e,s_0\right)\qquad.
\label{param3b}
\end{eqnarray}
It can be checked that this vector is invariant in the transformation
$(\Omega,\omega)\longrightarrow(\Omega+\pi,\omega+\pi)$. This will be our 
parameter vector for a second version of the MCMC code. The new Jacobian
 of the transformation from 
$(\ln q,e,\sin i,\Omega,\omega,t_p)$ to $\vec{w_3}$ reads now
\begin{equation}
J_3=q^2\sin^2i\sin^2\omega{\sqrt{1-\sin^2i\sin^2\omega}}\,\frac{\rd s}{\rd t}
\qquad.
\end{equation}
This new version succeeded in reaching convergence for \pz.
\subsection{Implementation}
The two versions of our code have been written in \textsc{Fortran 90},
with an additional OPEN-MP parallel treatment of the computed Markov
chains. Our basic strategy is the same is in \citet{beu14}. We first
perform a least-square Levenberg-Marquardt fit of the orbit. This
takes only a few seconds to converge towards a local $\chi^2$
minimum. Of course, this fit is made starting from a rough orbit
guess. The same procedure is re-initiated many times varying the
starting orbit. This allows to probe the variety of local $\chi^2$
minima. In all cases described below, among various minima, a main one,
or a series of very similar ones was reached. This main minimum was
selected as a starting point for the MCMC chains. This procedure turns
out to speed up the convergence of the chains. This starting point is
marked as red bars and black stars in the resulting MCMC posterior
plots (see below). We also tried to run the MCMC starting from a
random guess instead, and checked that the same posterior
distributions were reached, but slower. We also checked in the
posterior $\chi^2$ distributions derived from the runs that in all
cases, the starting point initially derived with Levenberg-Marquardt
actually achieves the best $\chi^2$ in the distribution. Strictly
speaking, Levenberg-Marquardt works well to quickly derive the best
$\chi^2$ minimum. This shows that using other least-square fitting
algorithms like downhill simplex for instance would not lead to a
better result, as all these methods aim at finding a $\chi^2$ minimum,
possibly the best one. However, MCMC runs reveal afterwards that with
sparsely sampled orbits as we are dealing with here, the very best
$\chi^2$ minimum does not always correspond to a probability peak in
the posterior distributions. This intrinsic fact is independent from
the method used to get the minimum.

The implementation of the universal variable formalism described above
requires an efficient algorithm to compute the Stumpff functions. The
series (\ref{ck}) defining them efficiently converge only for
sufficiently small $x$. We use a reduction algorithm described in
\citet{dan87}, that makes use of the following set of formulas
\begin{eqnarray}  
c_0(4x)=2\left[c_0(x)\right]^2,& \, &
c_1(4x)=c_0(x)c_1(x),\label{qu1}\\
c_2(4x)=\frac{1}{2}\left[c_2(x)\right]^2,&\, &
c_3(4x)=\frac{1}{4}c_2(x)+\frac{1}{4}c_0(x)c_3(x)\;.\label{qu2}
\end{eqnarray}
Any input argument $x$ is first reduced by successive factors of 4
until it satisfies $|x|<0.1$. Then the series up to order 6 are used
to get $c_2(x)$ and $c_3(x)$ only. To compute $c_0$ and $c_1$, the
following relations are used
\begin{equation}
c_0(x)=1-xc_2(x),\qquad c_1(x)=1-xc_3(x)\quad.
\end{equation}
Equations~(\ref{qu1}) and (\ref{qu2}) are then applied recursively to
derive the Stumpff functions for the original argument
$x$. \citet{dan87} demonstrated the efficiency of this algorithm.

In the fitting routine, universal Kepler's equation~(\ref{kepu}) must
be solved numerically using a root-finding algorithm. We do it with
Newton's quartic method or with Laguerre-Conway's method
\citep{dan87}. To compute the derivatives of the Stumpff function, we
use the following relation
\begin{equation}
\frac{\rd c_n(x)}{\rd x}=\frac{1}{2x}\left(c_{n_1}(x)-nc_n(x)\right)\qquad,
\end{equation}
or equivalently, if we define $\phi_n(\alpha,s)=s^nc_n(\alpha s^2)$,
\begin{equation}
\frac{\partial\phi_n(\alpha,s)}{\partial s}=\phi_{n-1}(\alpha,s)\qquad.
\end{equation}
For the special case $n=0$, we have
\begin{equation}
\frac{\partial\phi_o(\alpha,s)}{\partial s}=-\alpha\phi_1(\alpha,s)\qquad.
\end{equation}
The same algorithm is implemented in the symplectic $N$-body
integrator \textsc{Swift} \citep{ld94} for high eccentricity
orbits. Its use turns out to be only a few times (3--4) more computing
time consuming than that of a standard Keplerian formalism based on
sine and cosine functions. But this is worth applying it in MCMC to
the case of very high eccentricity and open orbits, as the use of the
universal Kepler's equation (\ref{kepu}) eliminates the instabilities
due to the permanent switch between the various formulas
(\ref{keps}). Thus Markov chains converge more efficiently.
\section{Results for \fomb}
\fomb\ is known to have a very eccentric orbit, with an
eccentricity in any case $\ga 0.5$ and most probably around 0.8--0.9
\citep{beu14,pea15}. Whether it is actually bound to the central star
may be questionable, especially due to the very small coverage
fraction of its orbit. If bound, its orbital period is a matter of
hundreds of years if not more, while the four available astrometric
points span over a period of 8 years only. Therefore, as noted by
\citet{pea15}, what is measured is basically a projected position and
a projected velocity onto the sky plane, so that the $z$-coordinates
(i.e., along the line of sight) of the position an velocity are
unknown. As a matter of fact, \citet{pea15} use with these data a
simple sampling method drawing random $z$-coordinates for the position
on velocity. They find an eccentricity distribution very similar to
that derived in \citet{beu14} with MCMC. However, in both cases, the
orbit was supposed to be bound. The eccentricity distribution of
\citet{pea15} nevertheless extends up to $e=1$ (that of \citet{beu14}
stops at $e\simeq 0.98$), showing that unbound solutions could exist
as well. This justifies the use of our new code to check this
possibility.

\begin{figure*}
\includegraphics[width=\textwidth]{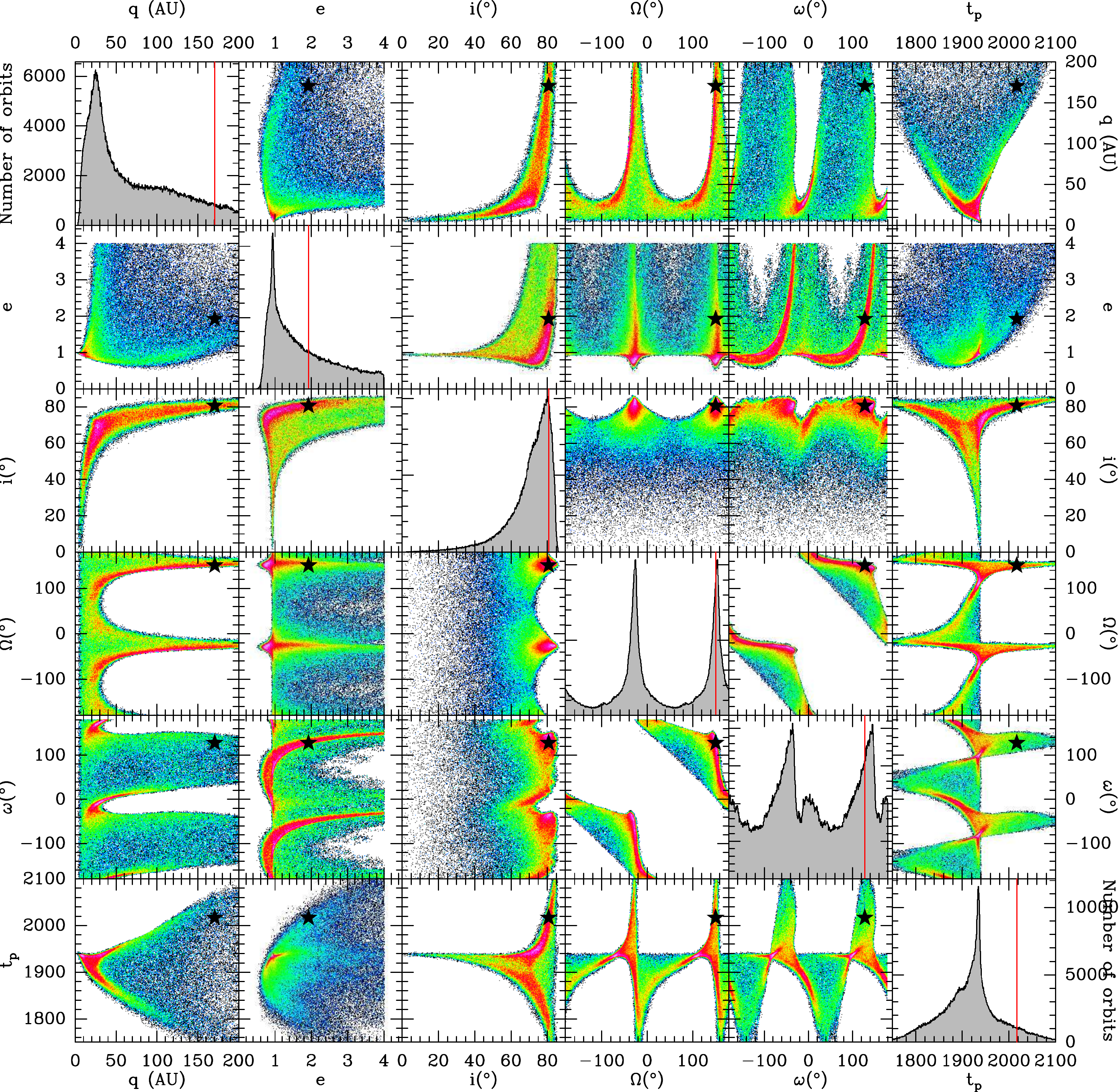}
\caption[]{Resulting MCMC posterior distribution of the six orbital
  elements ($q$, $e$, $i$, $\Omega$, $\omega$, $t_p$) of \fomb's
  orbit using the universal variable code. The diagonal diagrams show
  mono-dimensional probability distributions of the individual
  elements. The off-diagonal plots show bi-dimensional probability maps
  for the various couples of parameters. This illustrates the
  correlation between orbital elements. The logarithmic color scale in
  these plots is linked to the relative local density of orbital
  solutions. It is indicated on the side of Fig.~\ref{fombmcmc}. 
  In the diagonal histograms, the red bar indicates the location of the
  best $\chi^2$ solution obtained via standard least-square
  fitting. The location of this solution is marked with black stars in
  the off-diagonal plots.}
\label{mcmcmap_fomb}
\end{figure*}

The code in its first version (see above) was used with the available
astrometric data from \citet{kal13} and listed in
\citet{beu14}. Following the prescriptions by \citet{ford06}, 10
chains were run in parallel until the Gelman-Rubin parameters
$\hat{R}$ and $\hat{T}$ reach repeatedly convergence criteria for all
parameters in Eq.~(\ref{param2}), i.e., $\hat{R}<1.01$ and
$\hat{T}>1000$. We had already used the same procedure in
\citet{chau12} and \citet{beu14}. In \citet{beu14}, this convergence
criteria were reached after $6.2\times 10^7$ steps. Here it took
$4.25\times 10^8$ steps with the universal variable code, running on
the same data. This illustrates how the possibility for Markov chains
to extend in the unbound orbit domain increases the complexity of the
problem. We also had to fix an arbitrary upper limit
$e_\mathrm{max}=4$ for the eccentricity to ensure convergence. Setting
larger $e_\mathrm{max}$ values results in more steps needed to reach
convergence, but the assumed $e_\mathrm{max}=4$ upper limit as some 
physical justification. A large eccentricity means that \fomb\ is a
passing by object that is currently encountering a flyby in the 
Fomalhaut system. The eccentricity of a flyby orbit
cannot be arbitrarily large. For an hyperbolic orbit, the eccentricity
is directly linked to the relative velocity at infinity
$v_\mathrm{\infty}$ by the energy balance equation
\begin{equation}
\frac{1}{2}v_\mathrm{\infty}^2=\frac{GM(e-1)}{q}\qquad.
\label{vinfty}
\end{equation}
An upper limit to $v_\mathrm{\infty}$ can be given considering a
typical dispersion velocity in the solar neighborhood, i.e., $\sim
20\,\mathrm{km\,s}^{-1}$. Assuming $q=25\,$AU, i.e., the most probable
value for hyperbolic orbits in our distribution (see
Fig.~\ref{mcmcmap_fomb}), this immediately translates into an estimate of
an upper limit for the eccentricity $e_\mathrm{max}\simeq 4$. 

\begin{figure*}
\makebox[\textwidth]{
\includegraphics[width=0.33\textwidth]{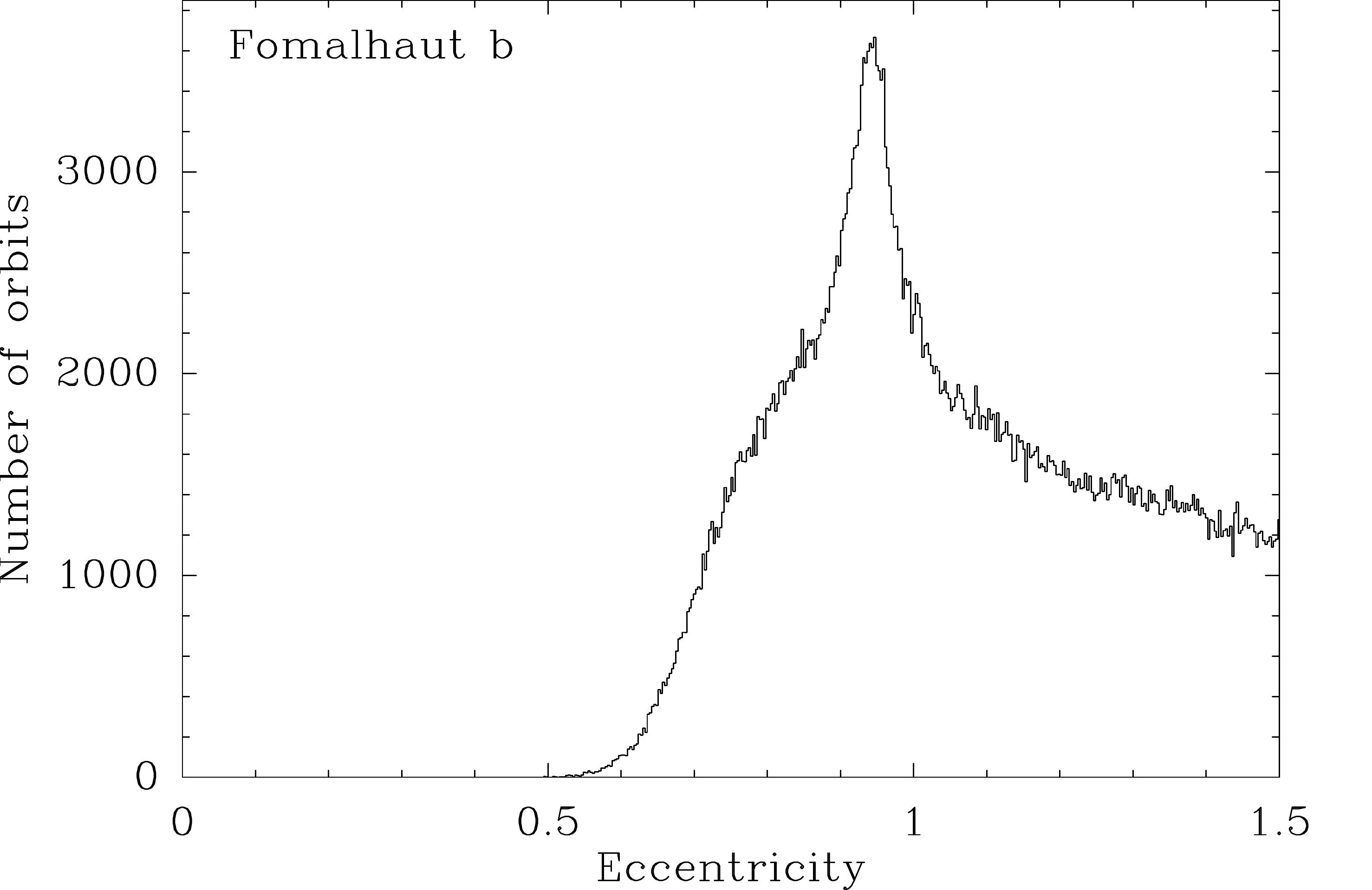} \hfil
\includegraphics[width=0.33\textwidth]{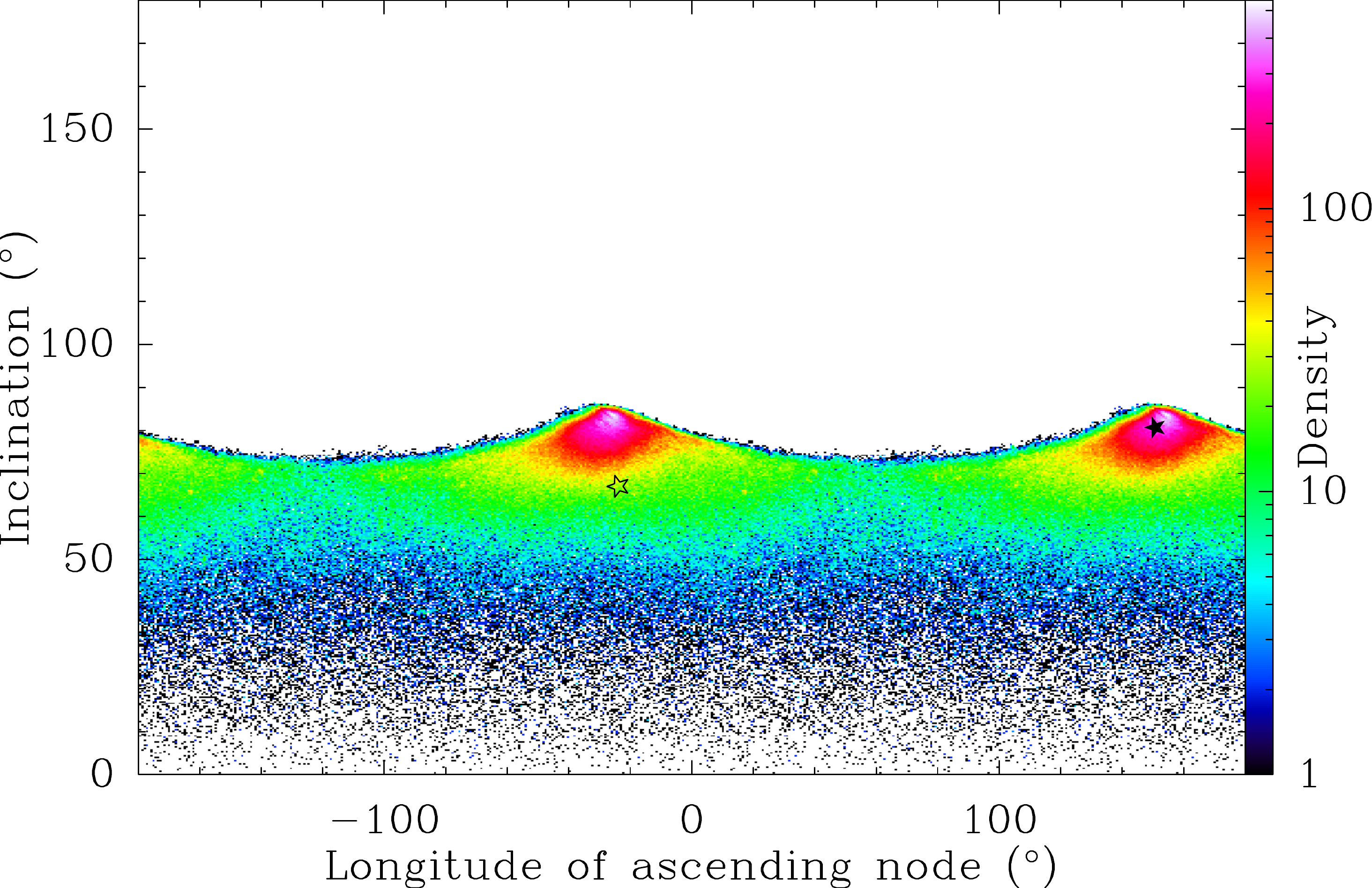} \hfil
\parbox[b]{0.3\textwidth}{
\caption[]{Enlargement of eccentricity histogram (left) and density maps 
in ($\Omega,i$) space (right) out of Fig.~\ref{mcmcmap_fomb}. The color 
scale appearing on the right side of the right plot map holds for all 
similar plots in Fig.~\ref{mcmcmap_fomb}. The eccentricity histogram is 
the same as in Fig.~\ref{mcmcmap_fomb}, except that is was truncated 
at $e=1.5$ to allow a better comparison with \citet{beu14}. In the  
$\Omega$--$i$ plot, the open star shows the estimated location of the 
disk plane \citep{kal05}, and the black star indicates the location 
of the best $\chi^2$ solution.}
\label{fombmcmc}}}
\end{figure*}

The global statistics of the posterior orbital distribution obtained
from the run is shown in Fig.~\ref{mcmcmap_fomb}, where distribution
histograms for all individual elements ($q$, $e$, $i$, $\Omega$,
$\omega$, $t_p$) as well as density maps for all combinations of them
are represented. Special enlargements of the plots concerning $e$  and $\Omega$
are shown in Fig.~\ref{fombmcmc}. In all histogram (mono-dimensional) plots, 
the red vertical bar superimposed to the
plot shows the corresponding value for the best fit (lowest $\chi^2$)
orbit obtained independently via a least-square Levenberg-Marquardt
procedure.  The same orbital solution is marked with black stars in 
all off-diagonal density plots combining two orbital parameters. 
As explained above, a least-square fit is initiated prior to 
launching MCMC. The resulting best-fit model is then used as a starting
point for the Markov chains, and posterior $\chi^2$ distributions show
that this solution actually achieves the minimum of the distribution.

We first compare these plots to the corresponding ones in
\citet{beu14}, where the fit was made over the same data set, but
limited to bound orbits only. The first striking fact is that the
eccentricity distribution extends now beyond $e=1$ well into the
unbound regime. This shows as suspected that unbound orbital solutions
for \fomb\ do exist. The best fit solution is itself an unbound 
orbit with $e\simeq 1.9$. We nevertheless note a strong peak in the 
distribution at $e\simeq 0.94$ that appears exactly at the same place as in
\citet{beu14}. This clearly shows that for such a weakly constrained
problem, MCMC is definitely superior to least-square.

In fact, the whole eccentricity distribution below $e\la 0.96$ exactly
matches the corresponding one in \citet{beu14}. This shows up in 
Fig.~\ref{fombmcmc} where the eccentricity histogram was 
intentionally cut at $e=1.5$
to permit a better comparison. This first validates
the present run (as the previous one was done with another code), and
second shows that the cutoff at $e\simeq 0.98$ that appeared in the
previous distribution was not physical, but rather due to the
intrinsic limitation of the code used. The tail of the distribution
extends now in the unbound regime up to the $e_\mathrm{max}=4$
limitation that was fixed in the run. The shape of this tail can be
fitted as with a $e^{-3/2}$ power law. We also note (Fig.~\ref{mcmcmap_fomb})
that the periastron distribution closely matches that of \citet{beu14}, while
extending further out towards larger values. This is clearly due to
the contribution of unbound orbits, as can be seen in the
$q$--$e$ probability map.

From this we can derive an estimate of the probability for \fomb's
orbit to be bound, just counting the number of bound orbits in the
whole set. We find $p_\mathrm{bound}=0.23$. This probability actually
depends on the assumed limitation $e_\mathrm{max}=4$. If we had let
the eccentricity to take larger values, the number of unbound orbits
in the whole set would have been larger, and subsequently
$p_\mathrm{bound}$ would have been smaller. It is nevertheless
  possible to estimate the ultimate $p_\mathrm{bound}$ value that
  would be derived if we put no upper limit on the
  eccentricity. Taking into account the fact that the tail of the
  posterior eccentricity distribution roughly falls off as $e^{-3/2}$,
  we can extrapolate the distribution up to infinity, integrate it and
  reintroduce the missing orbits corresponding to $e>4$ into the
  distribution. Our posterior sample of orbits contains $10^6$
  solutions. Extrapolating the distribution we can estimate that
  $\sim2.1\times10^5$ corresponding to $e>4$ are missing in our
  sample. This changes our probability estimate to
  $p_\mathrm{bound}=0.19$, which can be considered as a minimum
  value. However, as the $e_\mathrm{max}=4$ threshold results from a
physical consideration (see above), the first derived
$p_\mathrm{bound}$ value can be regarded as robust. Note also
  that it is not very much above the minimum value. This shows that
  the contribution of very high eccentricity solutions is minor.

This probability is however just derived from a pure mathematical
analysis without any likelihood consideration. Flybys are rare but not
necessarily improbable \citep[see example in][]{rec09}. Looking now at
the distributions of the other orbital elements, we see in 
Fig.~\ref{fombmcmc} that the
location of the orbit in $(\Omega,i)$ still closely matches that of
the observed dust disk (the white star in the plot) \citep{kal05},
like in \citet{beu14}. In other works, there is still a strong
suspicion of near-coplanarity between \fomb\ and the dust
disk. This clearly favours a bound configuration rather than a flyby
that would have no reason for being coplanar. Another possibility is
that \fomb\ is just getting ejected today from the system. That
last configuration nevertheless appears improbable regardless to the
timescale of the ejection ($\sim 1000$\,yrs) compared to the age of
the system \citep[440\,Myrs][]{mam12}. To conclude, these plausibility
considerations combined with our $p_\mathrm{bound}\simeq 0.23$
estimate and the clear peak of the eccentricity distribution at
$e=0.94$ enable us to stress that \fomb\ is probably bound to the
central star.

The situation is less clear with the argument of periastron. In the
$\omega$--$e$ plot (Fig.~\ref{mcmcmap_fomb}), we see that depending on
whether $e<1$ or $e>1$, the solutions exhibit different $\omega$
values. In \citet{beu14}, we noted that the observed dust disk
corresponds to $\omega_\mathrm{disk}=-148.9\degr$. This sill roughly
matches the $\omega$ values for bound orbits, i.e., bound orbits are
still apsidally aligned with the disk with a few tens of degrees.

\begin{figure}
\includegraphics[width=\columnwidth]{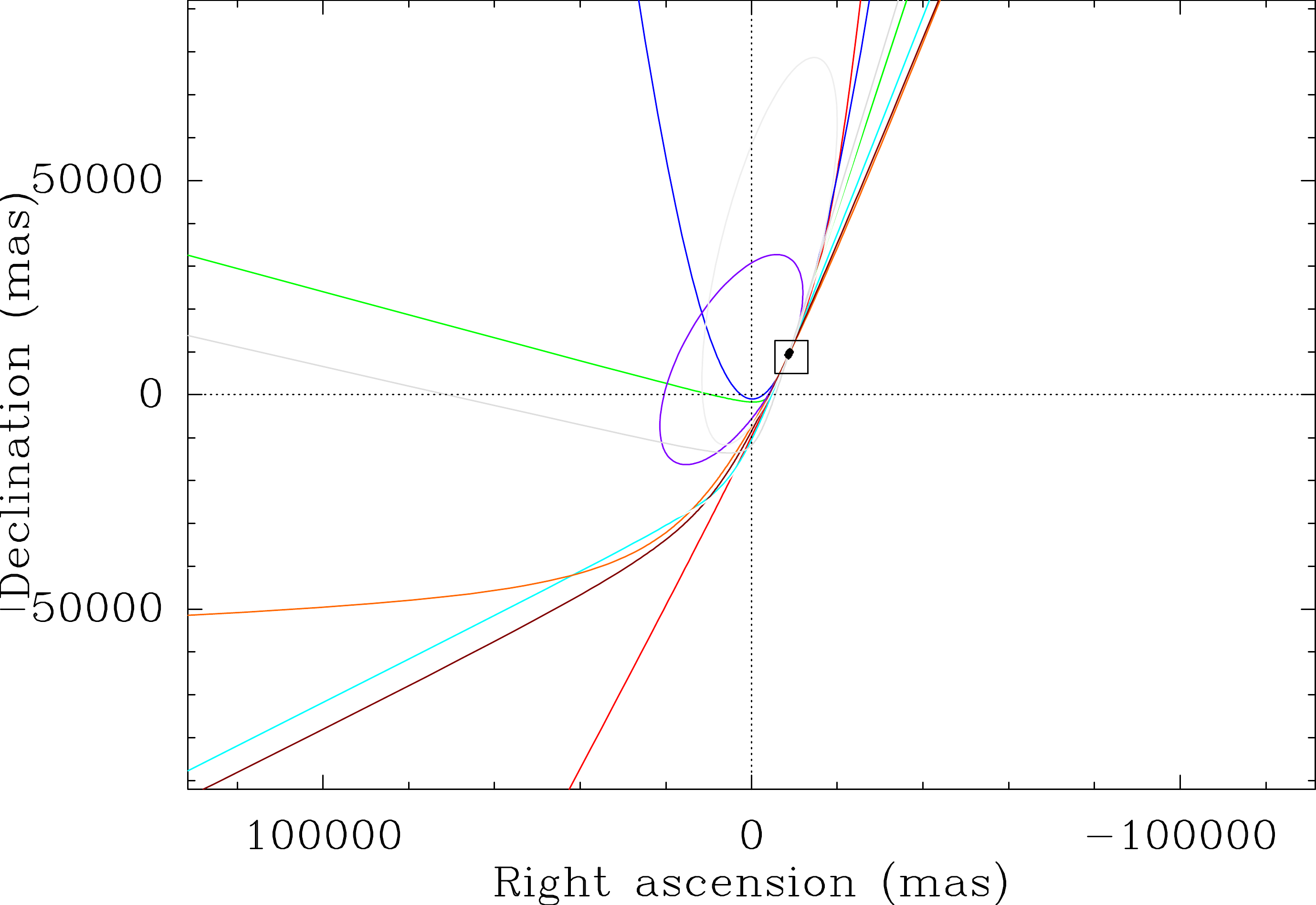}
\caption[]{Examples of orbital solutions for \fomb, in
  projection onto the sky plane. The star is at the center of the
  plot. The black square denotes the location of the observed
  astrometric points. Various colours are given to the orbits to 
  allow to distinguish them easily from each other}
\label{fomborbs}
\end{figure}
The distribution of the time of periastron passage ($t_p$) is similar
to that of\citet{beu14}, expect that we have here an additional tail
after 1950 that corresponds to unbound orbits. Figure~\ref{fomborbs}
shows finally a few orbital solution in projection onto the sky plane,
with bound and unbound configurations. We note that all solutions fit
the observed positions, while assuming very different shapes. We
clearly see here the effect of the bad observational orbital coverage.
\section{PZ Tel B}
\subsection{Results}
\begin{table*}
\caption[]{Summary of astrometric positions of \pz\ relative to PZ Tel 
as observed in recent years}
\label{pzteldata}
\begin{tabular*}{\textwidth}{@{\excs}llllll}
\hline\noalign{\smallskip}
Obs. date & Declination ($x$) & Right ascension ($y$) & Separation & Position Angle & Reference\\
 & (mas) & (mas) & (mas) & (mas) &\\
\noalign{\smallskip}\hline\noalign{\smallskip}
Jun. 13, 2007 & $121.26\pm1.20$ & $225.01\pm2.20$ & $255.6\pm2.5$ &
$61.68\pm0.6$ & \citet{mug12}\\
Apr. 11, 2009 & $169.96\pm8.57$ & $282.87\pm8.57$ & $330.0\pm10.$ & 
$59.0\pm1.0$ & \citet{bil10}\\
Sep. 28, 2009 & $165.65\pm0.59$ & $293.02\pm1.05$ & $336.6\pm1.2$ &
$60.52\pm0.22$ & \citet{mug12}\\
May 07, 2010  & $175.52\pm0.60$ & $308.23\pm1.04$ & $354.7\pm1.2$ &
$60.34\pm0.21$ & \citet{mug12}\\
May 05, 2010  & $183.25\pm1.53$ & $309.87\pm2.58$ & $360.0\pm3.0$ &
$59.4\pm0.5$ & \citet{bil10}\\
Sep. 26, 2010 & $186.90\pm4.10$ & $313.52\pm6.87$ & $365.0\pm8.0$ &
$59.2\pm0.8$ & This work\\
Oct. 28, 2010 & $185.15\pm0.55$ & $319.53\pm0.95$ & $369.3\pm1.1$ &
$59.91\pm0.18$ & \citet{mug12}\\
Mar. 25, 2011 & $192.02\pm0.51$ & $330.46\pm0.87$ & $382.2\pm1.0$ &
$59.84\pm0.19$ & \citet{mug12}\\
May 03, 2011  & $194.61\pm0.99$ & $342.58\pm1.74$ & $394.0\pm2.0$ &
$60.4\pm0.2$ & This work\\
Jun. 06, 2011 & $195.97\pm0.25$ & $335.22\pm0.43$ & $388.3\pm0.5$ &
$59.69\pm0.1$ &  \citet{mug12}\\
Jun. 07, 2011 & $195.00\pm2.51$ & $337.75\pm4.33$ & $390.0\pm5.0$ &
$60.0\pm0.6$ & This work\\
Apr. 05, 2012 & $196.09\pm4.45$ & $345.19\pm7.83$ & $397.0\pm9.0$ &
$60.4\pm0.2$ & \citet{bil13}\\
Jun. 08, 2012  & $212.41\pm0.10$ & $361.75\pm0.13$ & $419.5\pm0.14$ & 
$59.58\pm0.22$ & \citet{gin14}\\
\noalign{\smallskip}\hline
\end{tabular*}
\end{table*}
\begin{figure*}
\includegraphics[width=\textwidth]{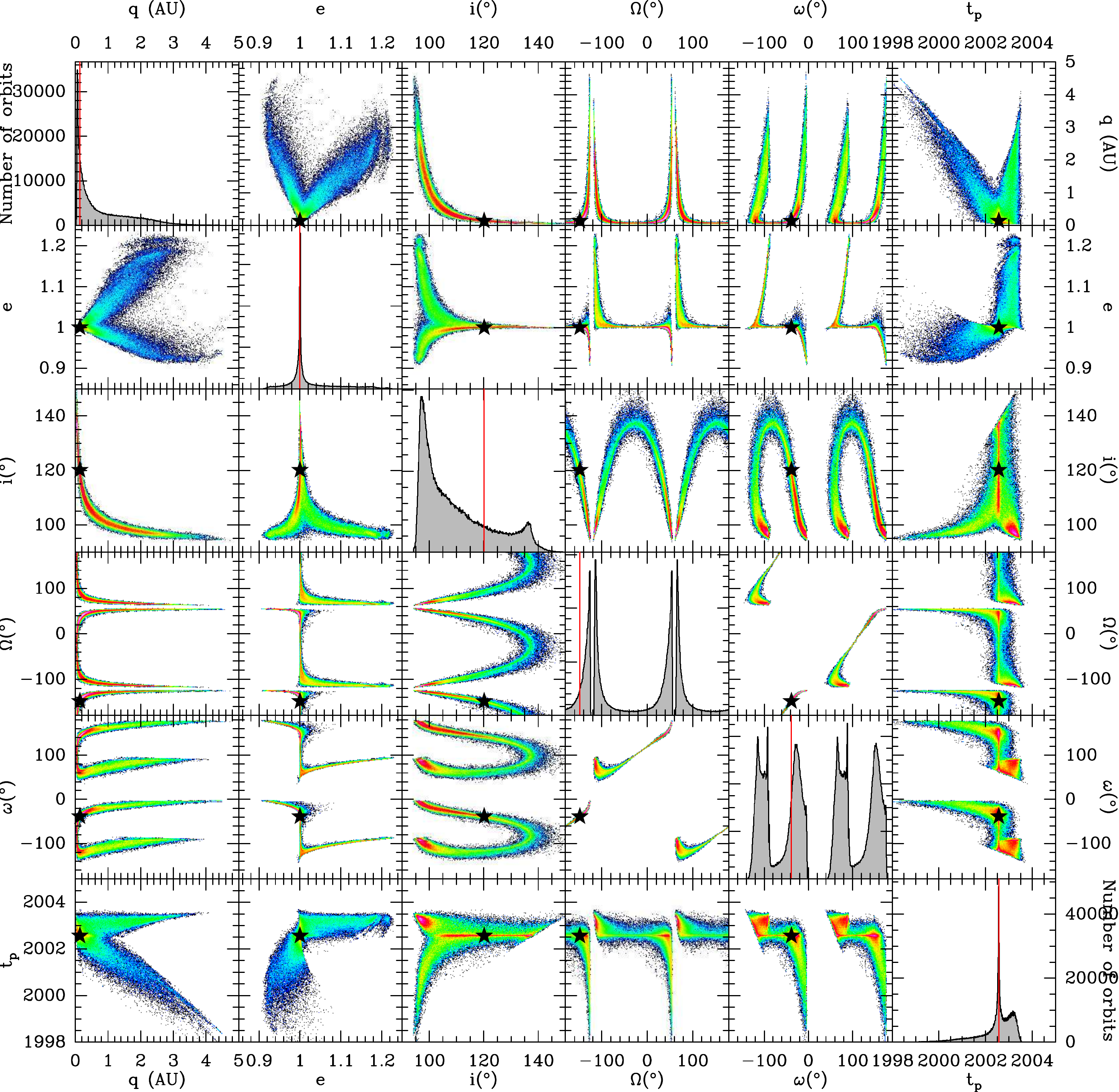}
\caption[]{Resulting MCMC posterior distribution of the six orbital
elements ($q$, $e$, $i$, $\Omega$, $\omega$, $t_p$) of \pz's
orbit using the universal variable code, presented in the same manner 
as for \fomb\ in Fig.~\ref{mcmcmap_fomb}. The plotting convention are 
identical.}
\label{mcmcmap_pztelb}
\end{figure*}
\begin{figure*}
\makebox[\textwidth]{
\includegraphics[width=0.33\textwidth]{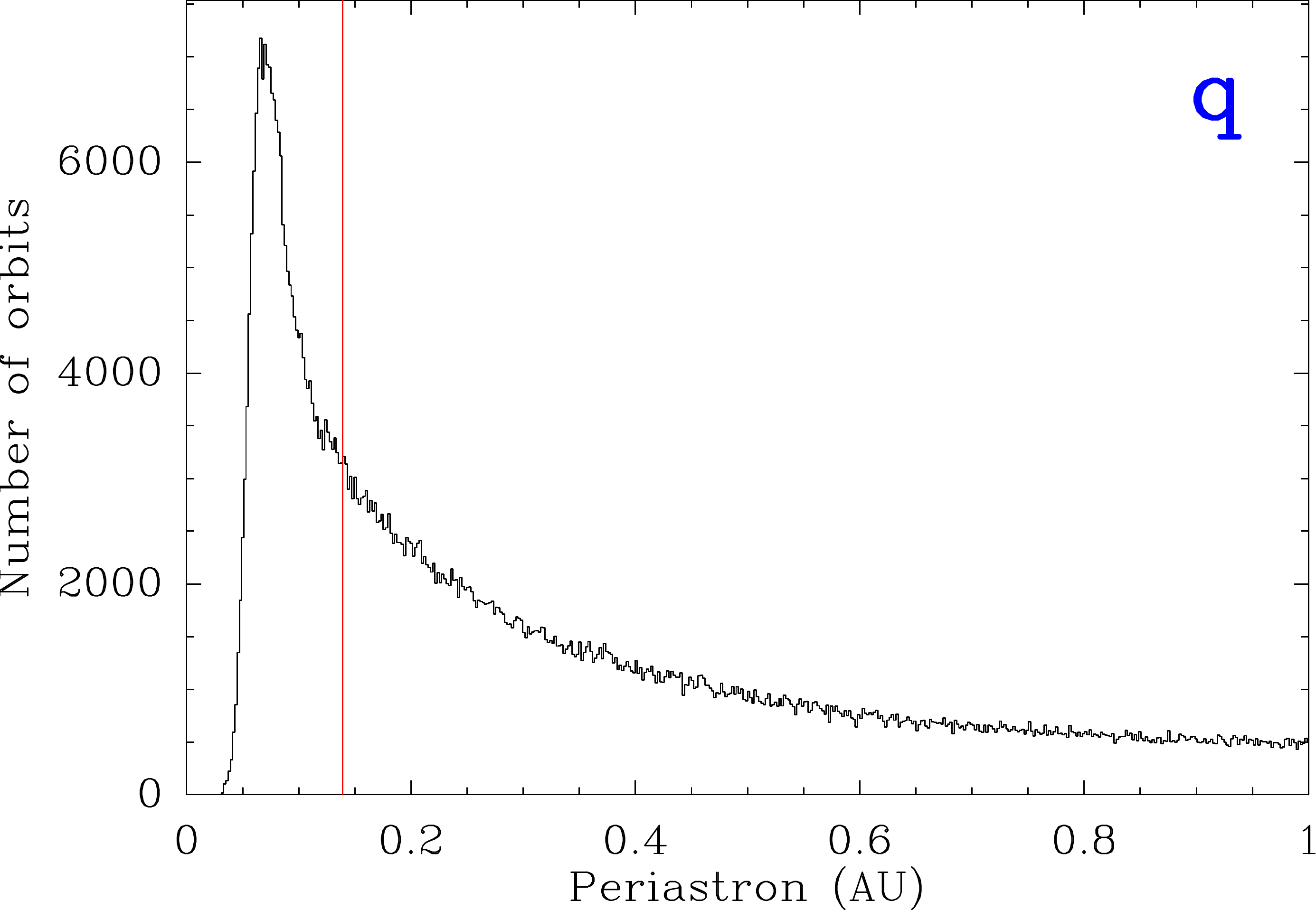} \hfil
\includegraphics[width=0.33\textwidth]{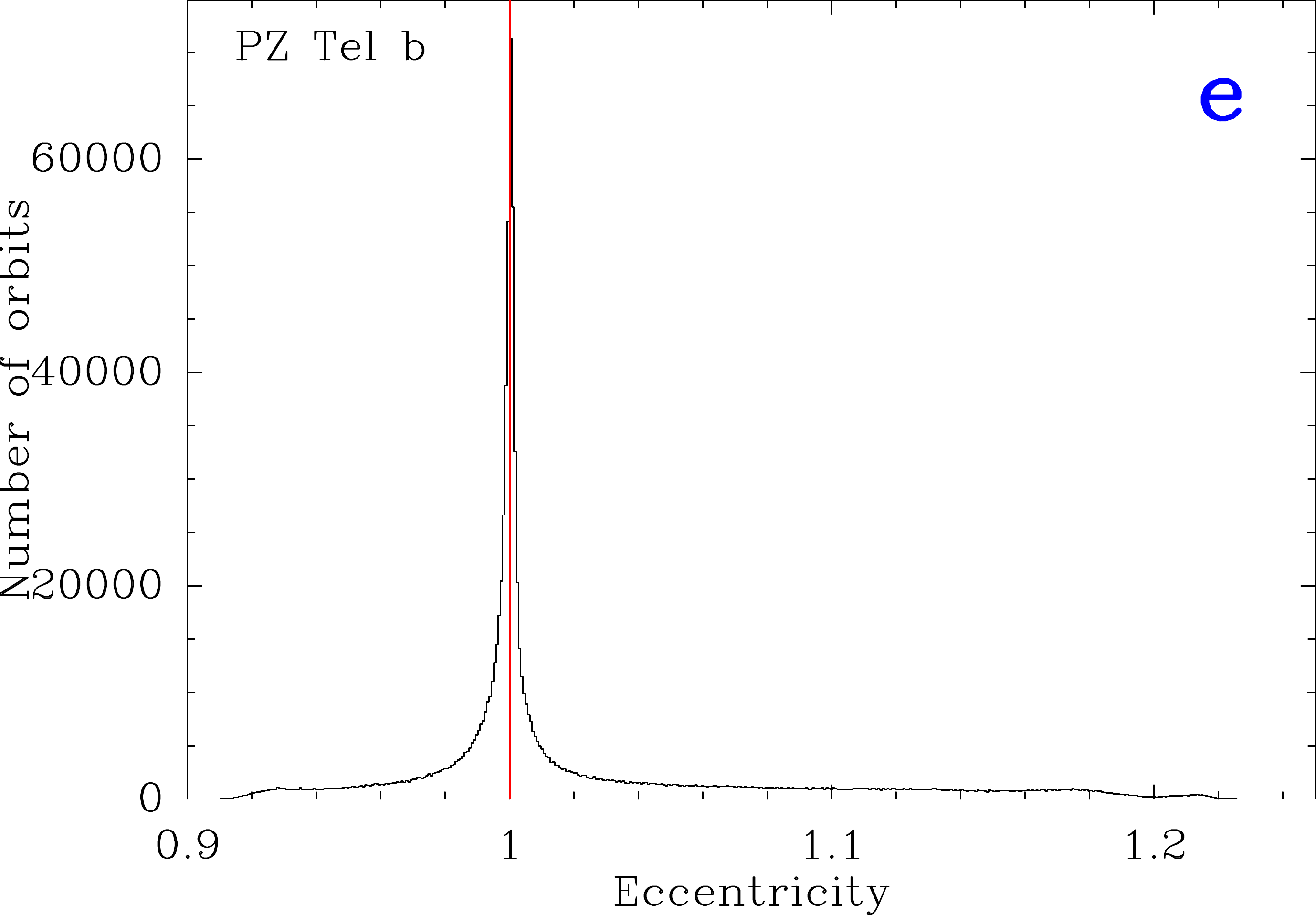} \hfil
\includegraphics[width=0.33\textwidth]{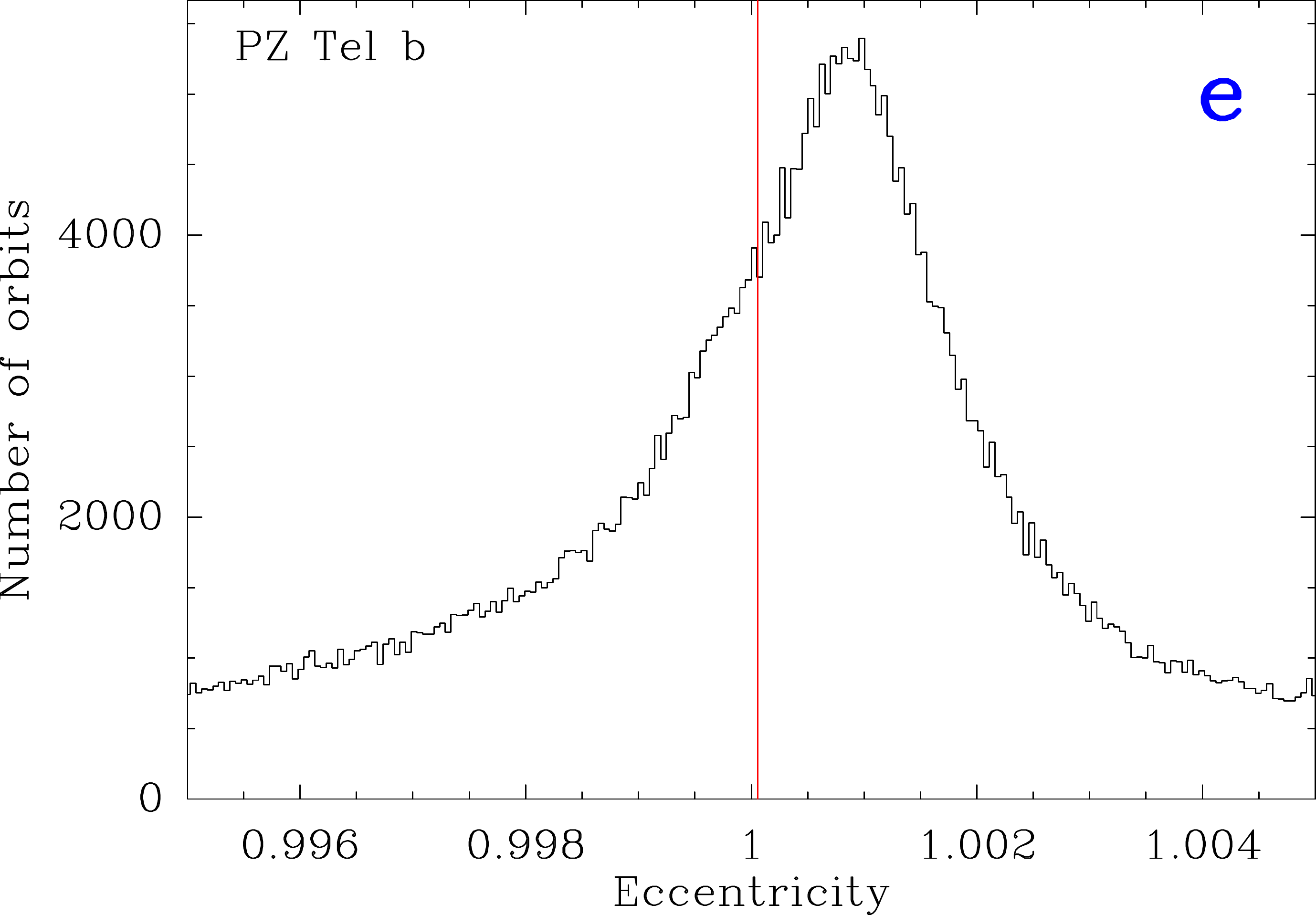}}
\caption[]{Enlargement of histograms and density maps out of 
Fig.~\ref{mcmcmap_pztelb} for the periastron ($q$) and eccentricity 
($e$) parameters. With respect to Fig.~\ref{mcmcmap_pztelb}, 
the periastron plot (left) was truncated at $q=1\,$AU to make 
the peak at $q=0.07$ appear. The right plot is a special zoom 
of the eccentricity distribution around $e=1$.} 
\label{pztelb_qe}
\end{figure*}
As mentioned above, orbital solutions for 
\pz\ \citep{bil10,mug12,gin14} all yields very eccentric orbits ($e>0.6$). 
These orbital determinations were
performed assuming a bound orbit, so that no test at $e\geq1$ was
done. This assumption is however questionable given the orbital
solutions found. We thus ran our universal variable MCMC code with the
available astrometric data of \pz\ (Table~\ref{pzteldata}). As
explained above (Sect.~2), we used the second version of the code with
parameter vector $\vec{w_3}$ (Eq.~\ref{param3b}) that ensures a better
convergence.

Even in that case, convergence was hard to reach. 10 chains were run
in parallel. After $1.5\times10^{10}$\,steps, the Gelman-Rubin
parameter $\hat{R}$ values for the 6 variables in $\vec{w_3}$ were
ranging between 1.006 and 1.019, while the $\hat{T}$ parameter values
were ranging between 260 and 800. The run was stopped there to save
computing time, as reaching the demanded criteria ($\hat{R}<1.01$ and
$\hat{T}>1000$ for all variables) would have demanded many more
steps. The $\hat{R}$ and $\hat{T}$ values reached at the stopping
point must nevertheless be considered as characteristic for an already
very good convergence, so that we may trust the resulting posterior
distribution. We checked indeed that posterior distributions that we
could derive stopping the computation earlier, i.e. at a point when
the $\hat{R}$ and $\hat{T}$ values were somewhat further away from
convergence criteria, did not show significant differences with those
presented below. Noticeably, nothing comparable in terms of
convergence criteria was reached using the first version of the code
using parameter vector $\vec{w_2}$ (Eq.~\ref{param2}).

The global statistics of the posterior orbital distribution obtained
from the run is shown in Fig.~\ref{mcmcmap_pztelb}, which was built
with the same plotting conventions as Fig.~\ref{mcmcmap_fomb} for
\fomb. In particular the red bar and the black star indicate the best
fit orbit obtained via Levenberg-Marquardt. Special
enlargements concerning the periastron and the eccentricity are shown
in Fig.~\ref{pztelb_qe}.

\begin{figure}
\includegraphics[width=\columnwidth]{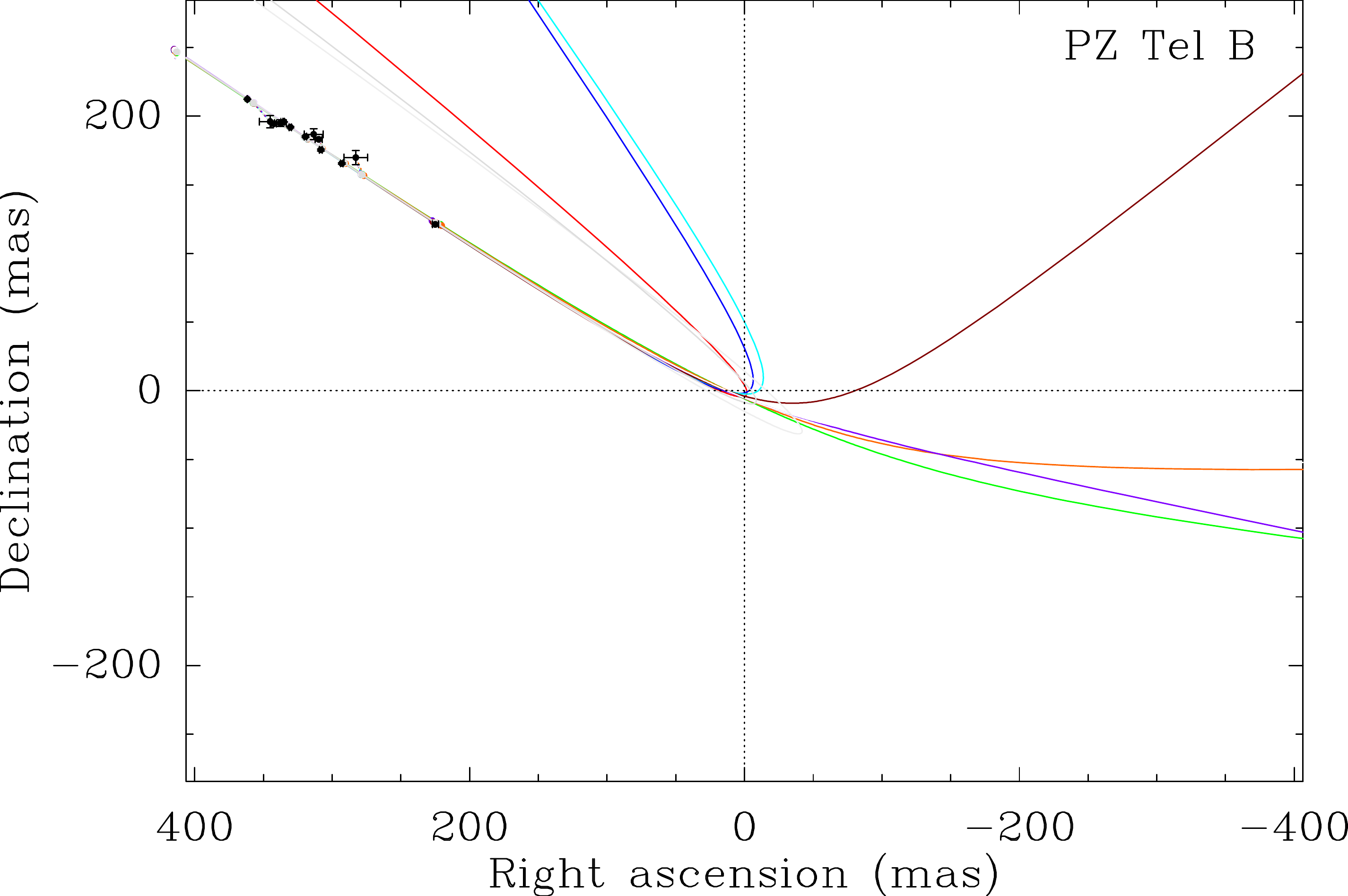}
\caption[]{Examples of orbital solutions for \pz, in
  projection onto the sky plane. The star is at the center of the
  plot. The data points appear in the upper left corner of the plot. 
As in Fig.~\ref{fomborbs}, colours are given to individual 
orbits to allow them to be distinguished from each other.}
\label{pztelb-orbits}
\end{figure}

The most striking feature that shows up is the eccentricity
distribution. As expected, \pz's orbit appears extremely
eccentric, but the eccentricity distribution is drastically different
from that of \fomb. As pointed out by \citet{pea15}, the
temporal coverage of \fomb's orbit is so small that what is
measured is basically a projected position and a projected velocity,
with no information about position and velocity along the line of
sight. Consequently, solutions with arbitrarily high eccentricities
are mathematically possible. This is not the case for \pz.
Table~\ref{pzteldata} shows that the motion followed over 5
years is quasi linear, but with a separation with the central star
that increased by more than 60\%. Even if the orbital coverage is
still small (see Fig.~\ref{pztelb-orbits}), this is more than just a
projected position and project velocity measurement. Consequently, no
tail extending to arbitrarily large values is obtained with MCMC. All
solutions naturally concentrate in the range $0.91<e<1.23$. The
eccentricity distribution appears extremely concentrated around
$e=1$. The enlargement in Fig.~\ref{pztelb_qe} close to $e=1$ reveals a
peak near $e=1.00$. The median of the distribution is at $e=1.001275$;
67\%\ and 95\%\ confidence levels are $0.965<e<1.024$ and
$0.906<e<1.157$ respectively. We compare the eccentricity distribution 
to that recently found by \citet{gin14}, who derived
$0.622<e<0.9991$ using a LSMC (Least-squares Monte-Carlo) approach, bur
restricted to bound orbits. Of course our distribution now extends
beyond $e=1$, but our lower bound ($e=0.91$) is significantly larger
than theirs. This is due to our additional data points rather than to
the method used.

The periastron distribution shows a
sharp peak around $q=0.07\,$AU. The $q$--$e$ map
(Fig.~\ref{mcmcmap_pztelb}) reveals actually two branches of solutions,
one with bound solution and one with unbound solutions. But most
solutions concentrate close to $e=1$ and $q=0.07\,$AU. The inclination
shows a peak at $i=98\degr$. This corresponds to a nearly edge-on
configuration, and could explain the quasi-linear motion. But
solutions up to $i=150\degr$ are also possible. The $i$--$e$ map shows
that the larger inclination solutions actually corresponds to those
with $e\simeq 1$. All solutions have $i>90\degr$, showing that the
orbit is viewed in a retrograde configuration from the Earth. The
distribution of the time of periastron $(t_p)$ exhibits a very sharp
peak in 2002.5 that corresponds also to orbits with $e\simeq 1$.
\begin{figure*}
\makebox[\textwidth]{
\includegraphics[width=0.49\textwidth]{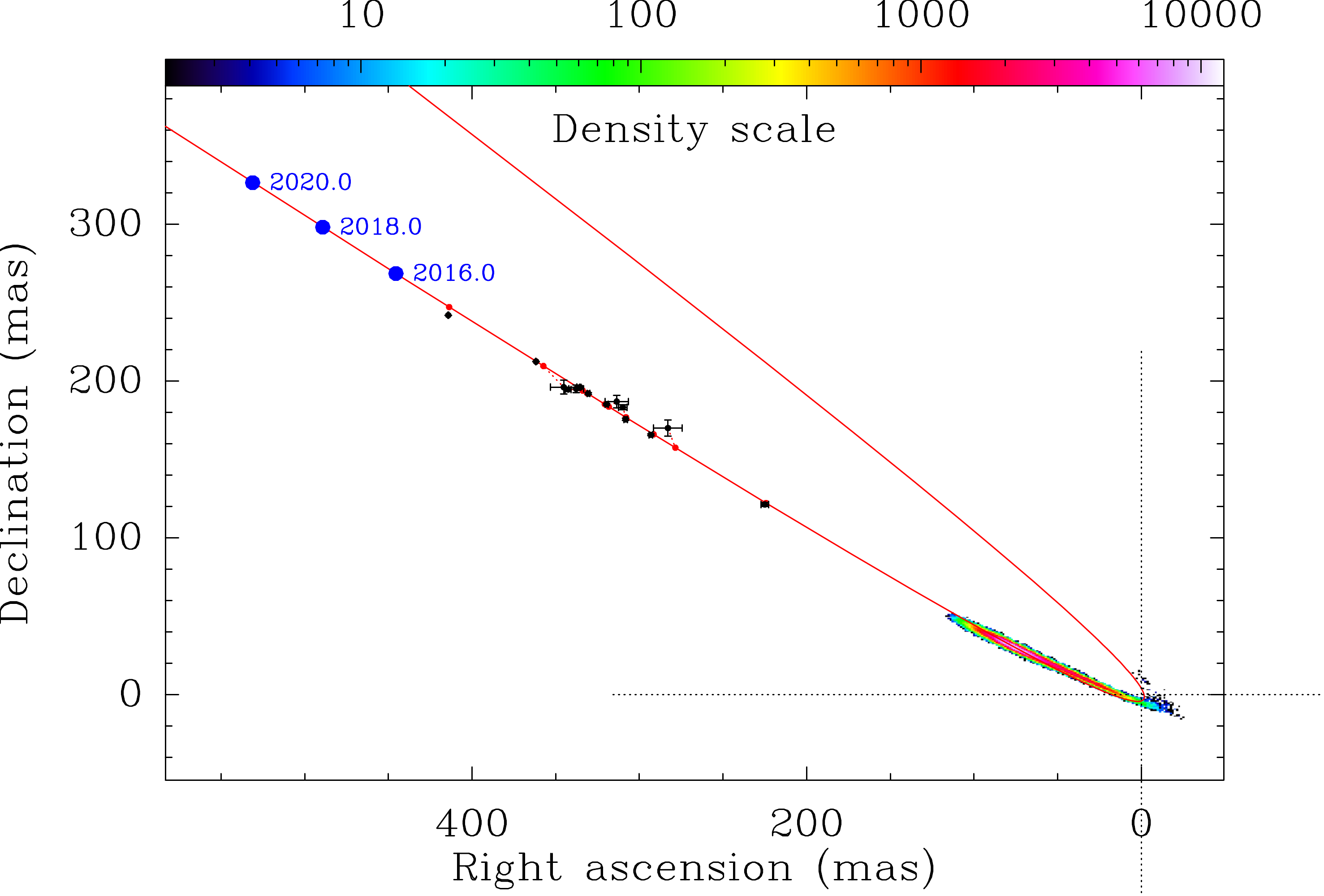}\hfil
\includegraphics[width=0.49\textwidth]{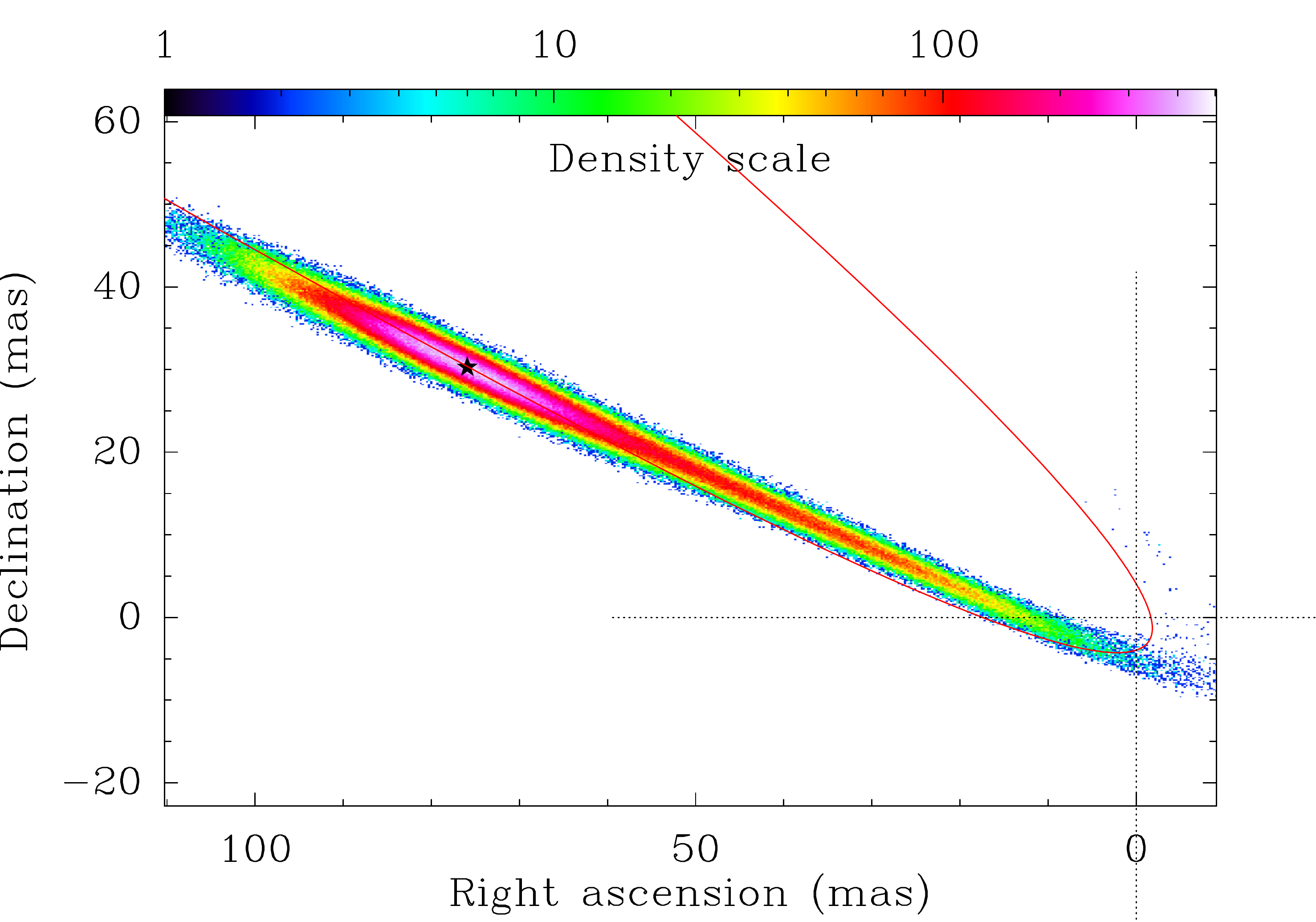}}
\caption[]{Density map of extrapolated projected position of \pz\ on 
Jul. 22, 2003, superimposed to the best $\chi^2$ orbit. 
\textbf{Left plot:} Large scale map showing the data points, 
the orbit, the predicted position for that orbit on 2016.0, 
2018.0, 2010.0 (blue points), and the cloud of positions on 
Jul. 22, 2003, computed for all solutions out or our posterior 
sample (shortly after periastron); \textbf{Right plot:} Enlargement 
of the periastron region with the cloud of projected positions. 
The peak of the distribution is marked with a black star.}
\label{pztelb-2003}
\end{figure*}

Figure~\ref{pztelb-orbits} shows a few orbital solutions in projection
onto the sky plane. We see that the solutions actually fit the data
points, but that they all have a much smaller
periastron. Figure~\ref{pztelb-2003} shows the best $\chi^2$ orbit in
a similar way, but superimposed to a density map showing the predicted
projected positions as of Jul. 22, 2003 (2003.556). \citet{mas05}
report indeed a non-detection of \pz\ in a NACO image from that
day. They conclude that no giant planet was present at a separation
larger than 170\,mas from the star. We see on Fig.~\ref{pztelb-2003}
that at the corresponding epoch, we predict that \pz\ was much
closer to the star, in most cases short after periastron, but in any
case closer to the star than 170\,mas. Our prediction is therefore in
agreement with the non-detection by \citet{mas05}.
\subsection{Discussion}
The exact nature of \pz's orbit around PZ~Tel is still
controversial. Our orbital analysis shows that both bound or unbound
solutions are valid, but that the eccentricity distribution is
strongly peaked around $e=1$. If \pz\ is unbound to PZ~Tel, it could
be a passing by object (a flyby). This is hard to believe given the
very small periastron value, and also considering the fact that both
the star and the companion seem to be young objects \citep{schm14}.
This cannot however be ruled out mathematically. From
Eq.~(\ref{vinfty}) with $M=1.25\,\msun$ and
$v_\infty=20\,$km\,s$^{-1}$, we derive $e=1.012$ for $q=0.07\,$AU and
$e=1.181$ for $q=1\,$AU, which is still in the tail of the
eccentricity distribution, although not in the main peak. We
nevertheless think this possibility to be very unlikely.

Still if \pz\ is unbound, it could be an escaping companion that was
recently ejected by some gravitational perturbation. Two problems
arise with this hypothesis. First, the ejection must have occurred
\emph{very} recently (a few years ago). Given the age of the star
\citep[23 Myr][]{mam14,bin14,malo14}, the probability of witnessing
such an event right today is very low, about $\sim 10^{-6}$ if we
consider the timescale of the ejection ($\sim 10\,$yrs) and the fact
that such an ejection should not occur more than a few times in the
history of the system. Second, to efficiently perturb a $40\,\mjup$
companion, an additional object of comparable mass (at least) is
required. As of yet, no such additional companion has been detected.

So, \pz\ is presumably a bound companion. Then one must 
explain its extremely high eccentricity. It could result from secular 
perturbation processes such as the Kozai-Lidov mechanism 
\citep{koz62,kry99,ford00}. This is a likely mechanism for generating 
very high eccentricities. But here again, given the fairly high mass of \pz, 
this would require a more massive outer companion that would probably 
have been already discovered.
\section{Hidden inner companions ?}
\subsection{\fomb}
\citet{pea14} argue that imaged substellar companions that appear very
eccentric with a first order orbital fit could actually be much less
eccentric due to the presence of an unseen inner companion. The reason
is that the measured astrometry is necessarily relative to the central
star, while in the presence of a massive enough inner companion, the
Keplerian motion of the imaged body should be considered around the
center of mass of the system. \citet{pea14} develop a full analytic
study showing how the presence of such an unseen companion could
artificially enhance the fitted eccentricity.

\citet{pea14} present a detailed study dedicated to the case of
\fomb, concluding that in any case, \fomb\ must be
significantly eccentric. According to them, in the best realistic
configuration, a $\sim 12\mjup$ companion orbiting Fomalhaut at 10\,AU
could account for a $\sim 10\%$ overestimate of \fomb's orbital
eccentricity. This would for instance shift the eccentricity peak from
$e=0.94$ to $e=0.85$. This possibility cannot be ruled out until the
inner configuration of Fomalhaut's planetary system remains
unconstrained. It would also be compatible with the scenario outlined
by \citet{far15} to explain the origin of \fomb's
eccentricity. According to this model, \fomb\ should have
formerly resided at $\sim 60\,$AU in the 5:2 mean-motion resonance
with another Jupiter-sized planet (termed Fomalhaut~c) 
located at $\sim 100\,$AU. Then, due
to the resonant action, its eccentricity would have increased, and it
would have been ejected towards its present-day orbit. This is still
compatible with the hypothetical presence of another massive planet
orbiting inside at 10\,AU. The only difficulty is that with $e=0.85$,
\fomb's periastron is still as low as $\sim 18\,$AU, which is
still close enough to 10\,AU to raise the question of its orbital
stability versus perturbations by the hidden companion. However,
according to \citet{far15}'s scenario, \fomb's orbit must
already cross today that of the putative Fomalhaut~c planet orbiting
at 100\,AU. So in any case, \fomb\ must lie today on a metastable
orbit. Adding another massive body deep inside the system does not
change this conclusion. Consequently, the presence of an additional
massive planet orbiting Fomalhaut at 10\,AU that would artificially
enhance \fomb's eccentricity by $\sim 10\%$ cannot be ruled out,
as being still compatible with all observational constraints. Moreover
it does not affect the dynamical scenario of \citet{far15}.
\subsection{PZ Tel B}
The case of \pz\ is more complex. The main difference with
\fomb\ is that it is imaged over a more significant part of its
orbit. As noted by \citet{pea15}, the detected astrometric motion of
\fomb\ is basically compatible with a straight line at constant speed,
so that what is measured is not much more than a projected position 
and a projected
velocity. This is not the case for \pz, as the projected distance
to the star is already able to vary significantly over the observation
period (Table~\ref{pzteldata}). This is actually the reason why the
fitted eccentricity distribution does not extend towards arbitrarily
large values (Fig.~\ref{pztelb_qe}). Consequently, an analytic study
of the potential effect of an unseen companion on the fitted orbit is
less easy. \citet{pea14} nevertheless calculated that a companion at
least as massive as $130\,\mjup$ orbiting PZ~Tel at $5.5\,$AU is
required to mimic \pz's eccentricity. But recent imaging by
\citet{gin14} exclude companions more massive than $26\,\mjup$ at this
distance. 

However, as for \fomb, an unseen companion could account only
partially for \pz's eccentricity. We decided to perform an
automated search based on this idea. Our strategy is the following: We
arbitrarily fix the characteristics of an unseen companion (mass and
orbit) that we may call \pzc. Given these characteristics, we calculate 
at each time the
expected position of the center of mass of the system, and recompute
the astrometric positions of \pz\ relative to this center of
mass. Then we restart a least-square fit and check the eccentricity of
the best $\chi^2$ solution obtained. This process is then
automatically re-initiated many times, changing the characteristics of 
\pzc\ until a solution yielding a least-square fit
with the minimal eccentricity is found. Of course we do several attempts, 
varying the starting points. They showed that in any case, \pz's 
eccentricity of the best $\chi^2$ solution never gets below $\sim 0.7$. 

For these most favourable configuration, the MCMC fit is re-launched to 
derive the statistical distributions of orbits. We present here one of 
these runs.  
\begin{table}
\caption[]{Characteristics of a putative \pzc\ that leads 
a less eccentric solution for \pz}
\label{pztelc}
\begin{tabular*}{\columnwidth}{@{\excs}ll}
\hline\noalign{\smallskip}
Mass & $12.041\pm0.1\,\mjup$\\
Semi-major axis & $3.514\pm0.0004$\,AU\\
Eccentricity & $0.4691\pm0.03$\\
Inclination & $\sim 0\degr$ ($4.924\times10^{-6}\pm3,$degrees)\\
Argument of periastron & $30.170\pm 15\,$degrees\\
Longitude of ascending node & $81.660\pm5\,$degrees\\
Time of periastron passage & $1951.57\pm0.25\,$AD\\
\noalign{\smallskip}\hline
\end{tabular*}
\end{table}
\begin{table}
\caption[]{Astrometric positions of \pz\ relative to the 
center of mass of the PZ~Tel -- \pzc\ system, computed from the data of 
Table~\ref{pzteldata} and with \pzc's characteristics taken from 
Table~\ref{pztelc}} 
\label{pzteldatamodif}
\begin{tabular*}{\columnwidth}{@{\excs}lll}
\hline\noalign{\smallskip}
Obs. date & Declination ($x$) & Right ascension ($y$)\\
& (mas) & (mas)\\
\noalign{\smallskip}\hline\noalign{\smallskip}
Jun. 13, 2007 & $120.86\pm1.20$ & $225.83\pm2.20$\\
Apr. 11, 2009 & $169.36\pm8.57$ & $282.88\pm8.57$\\
Sep. 28, 2009 & $165.30\pm0.59$ & $292.76\pm1.05$\\
May 07, 2010  & $175.76\pm0.60$ & $307.99\pm1.04$\\
May 05, 2010  & $183.50\pm1.53$ & $309.64\pm2.58$\\
Sep. 26, 2010 & $187.34\pm4.10$ & $313.61\pm6.87$\\
Oct. 28, 2010 & $185.60\pm0.55$ & $319.70\pm0.95$\\
Mar. 25, 2011 & $192.43\pm0.51$ & $330.92\pm0.87$\\
May 03, 2011  & $194.99\pm0.99$ & $343.11\pm1.74$\\
Jun. 06, 2011 & $196.32\pm0.25$ & $335.80\pm0.43$\\
Jun. 07, 2011 & $195.35\pm2.51$ & $338.33\pm4.33$\\
Apr. 05, 2012 & $196.11\pm4.45$ & $346.04\pm7.83$\\
Jun. 08, 2012  & $212.36\pm0.10$ & $362.62\pm0.13$\\
\noalign{\smallskip}\hline
\end{tabular*}
\end{table}
\begin{figure*}
\makebox[\textwidth]{
\includegraphics[width=0.33\textwidth]{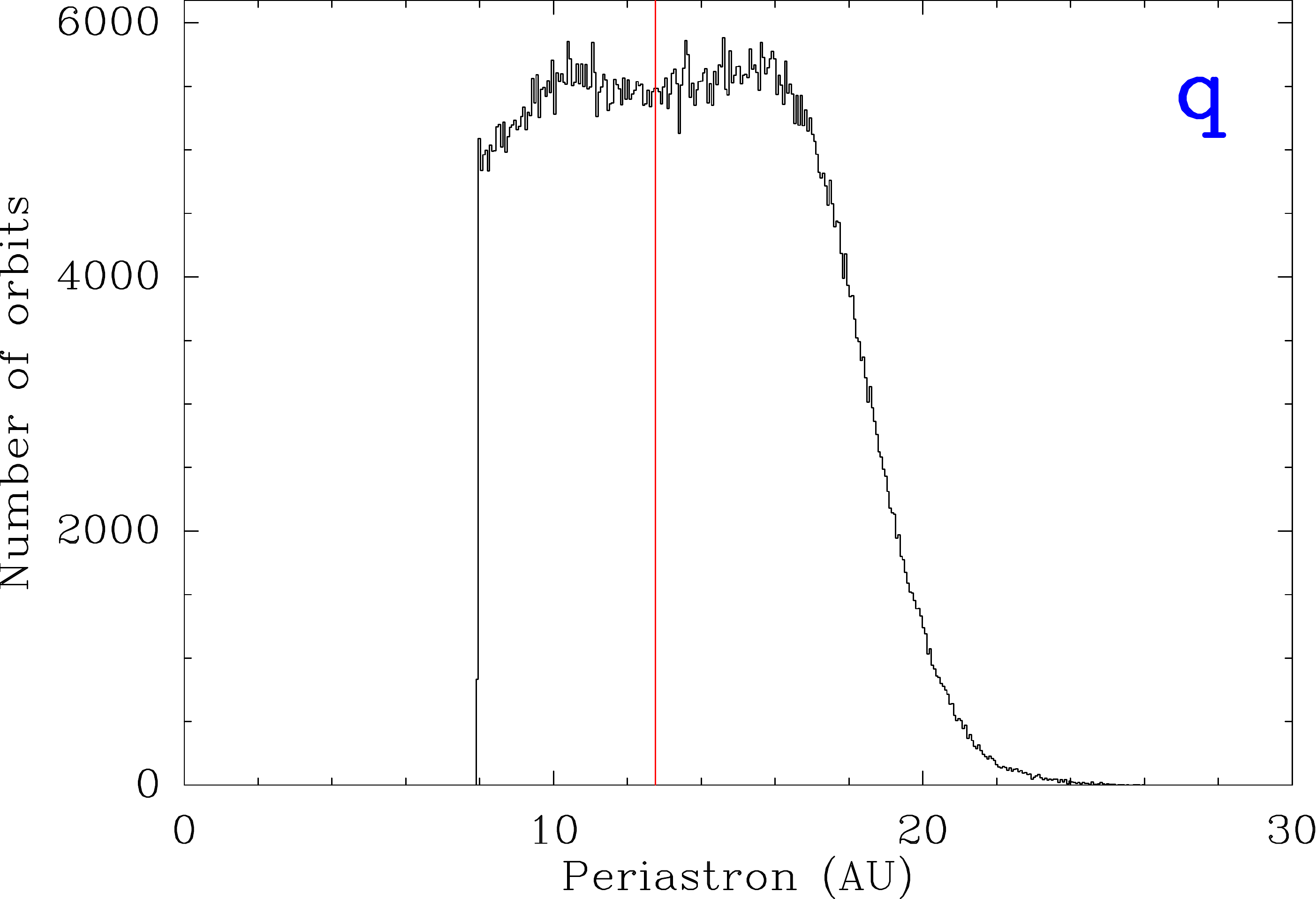} \hfil
\includegraphics[width=0.33\textwidth]{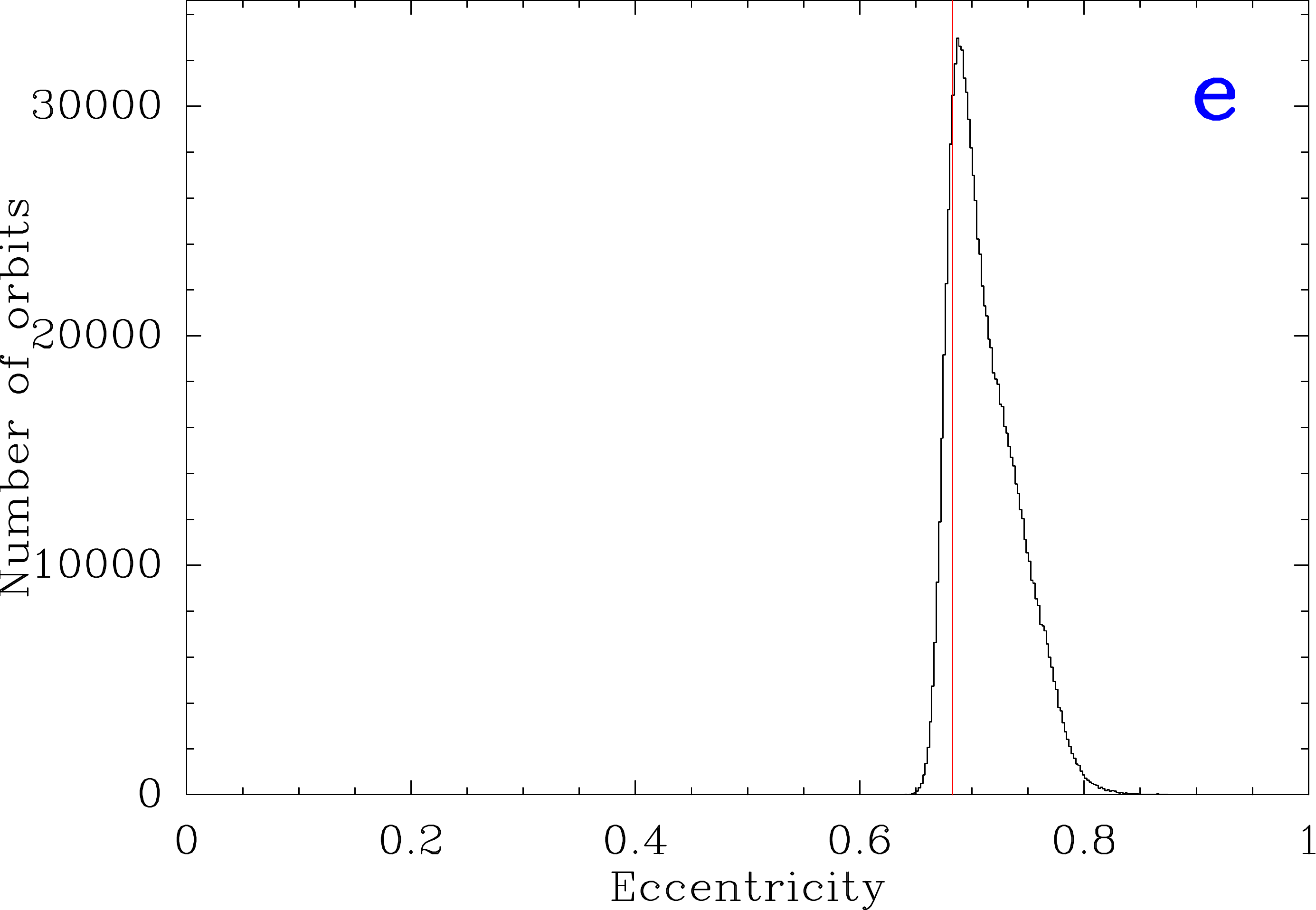} \hfil
\includegraphics[width=0.33\textwidth]{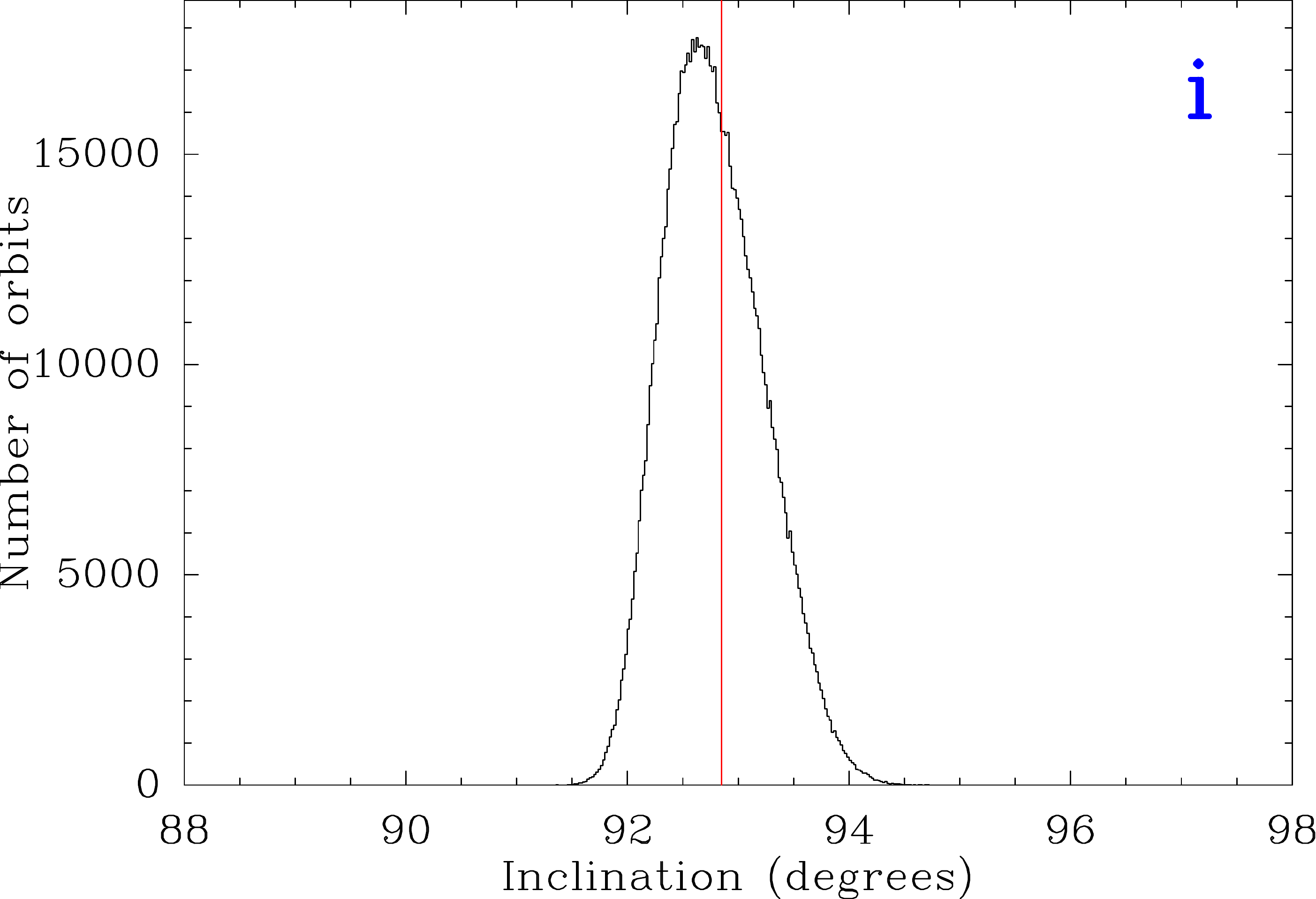}}
\caption[]{Resulting MCMC posterior distribution of the periastron
$q$ (left), the eccentricity $e$ (middle) and the inclination $i$ (right)
of \pz's orbit using the universal variable code, based on the modified
data from Table~\ref{pzteldatamodif}. The plotting conventions are the same 
as in Fig.~\ref{pztelb_qe}.}
\label{mcmcmap_pztelb_dk3}
\end{figure*}
The characteristics of the putative \pzc\ corresponding to this run
are listed in Table~\ref{pztelc} and the recomputed barycentric
astrometry of \pz\ is given in Table~\ref{pzteldatamodif}. The main 
characteristics of the 
result of the run (histograms of periastron, eccentricity and inclination) 
are shown in Fig.~\ref{mcmcmap_pztelb_dk3}. The first
thing we note is that the putative \pzc\ companion ($12\,\mjup$ at
3.5\,AU) is compatible with the current non-detection limits
\citep{gin14}. Second, the difference between the computed barycentric
astrometric data and the measured data is small, often within the
error bars of Table~\ref{pzteldata}. Nonetheless, the difference in
the orbital fit (Fig.~\ref{mcmcmap_pztelb_dk3}) is striking. \pz\
still appears eccentric, but its eccentricity is now confined between
0.65 and 0.8, the best $\chi^2$ orbit (the red bar in
the plots) having $e\simeq 0.68$. According to this analysis, \pz\
would clearly be a bound companion. Its periastron $q$ ranges between 8 and
24~AU, but it must be noted that all orbits with $q<8\,$AU have been
eliminated in the fitting procedure as being presumably highly unstable
versus gravitational perturbations from the hypothetical \pzc
companion. Letting the periastron distribution extend towards lower values
would generate solutions with larger eccentricities, but these would
probably not be physical.
\subsection{Dynamical analysis}
Figure~\ref{mcmcmap_pztelb_dk3} also reveals that \pz's orbital
inclination is very close to $90\degr$, meaning an almost edge-on
viewed orbit.  Simultaneously, \pzc's inclination appears extremely
low (Table~\ref{pztelc}), meaning a pole-on orbit. This allows to
question the dynamical stability of such a system. According to our
fit, \pz's periastron is most probably around $\sim 10\,$AU, which is
$\sim ~3$ times larger than the semi-major axis of \pzc's semi-major
axis. This is in principle marginally enough to ensure the dynamical
stability of the whole. But here both orbits are nearly
perpendicular. In this context, the inner orbit is likely to be
trapped in the Kozai-Lidov mechanism \citep{koz62,kry99,ford00}
characterized by huge eccentricity changes. This could trigger orbital
instability.

\begin{figure*}
\makebox[\textwidth]{
\includegraphics[width=0.33\textwidth]{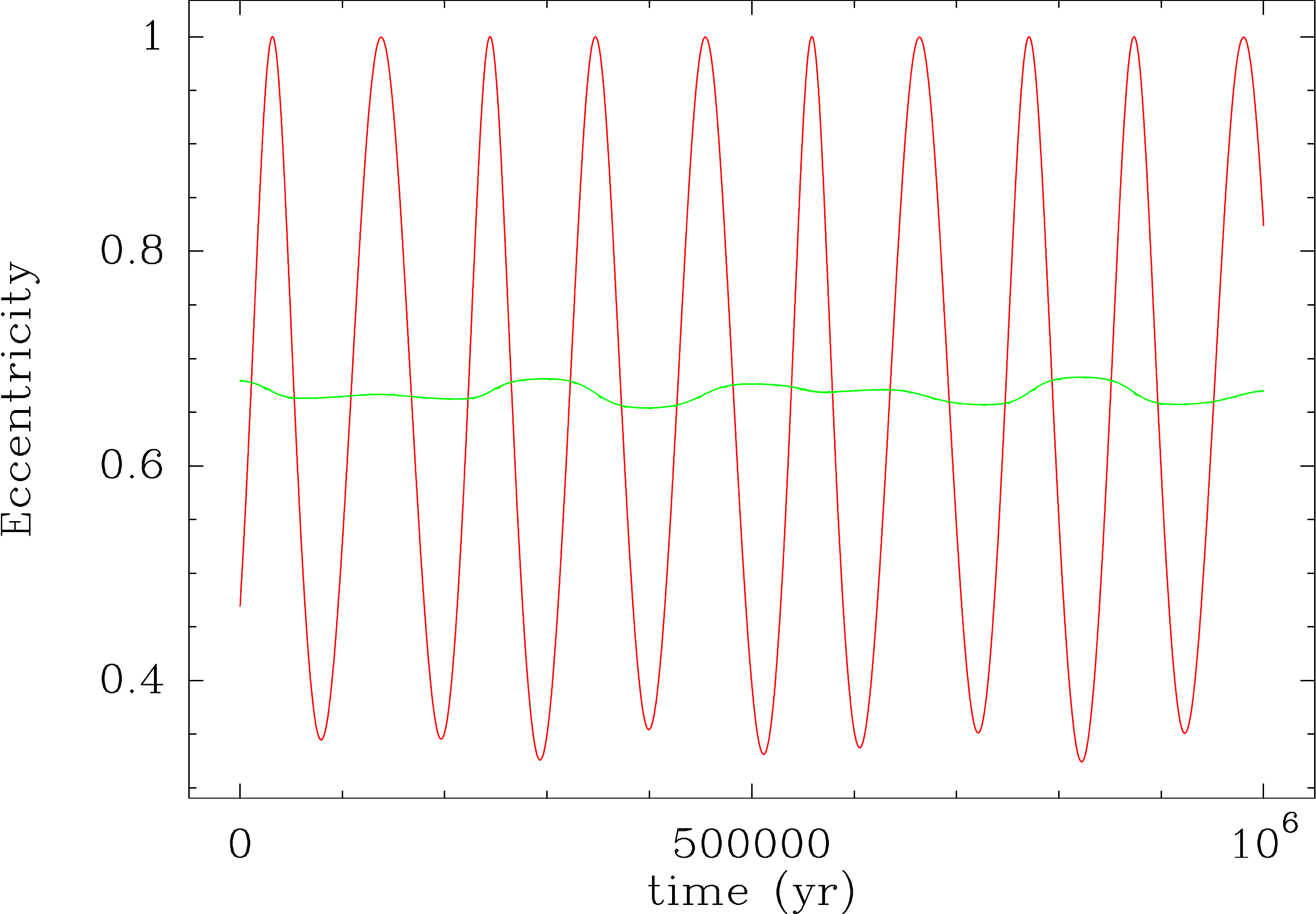} \hfil
\includegraphics[width=0.33\textwidth]{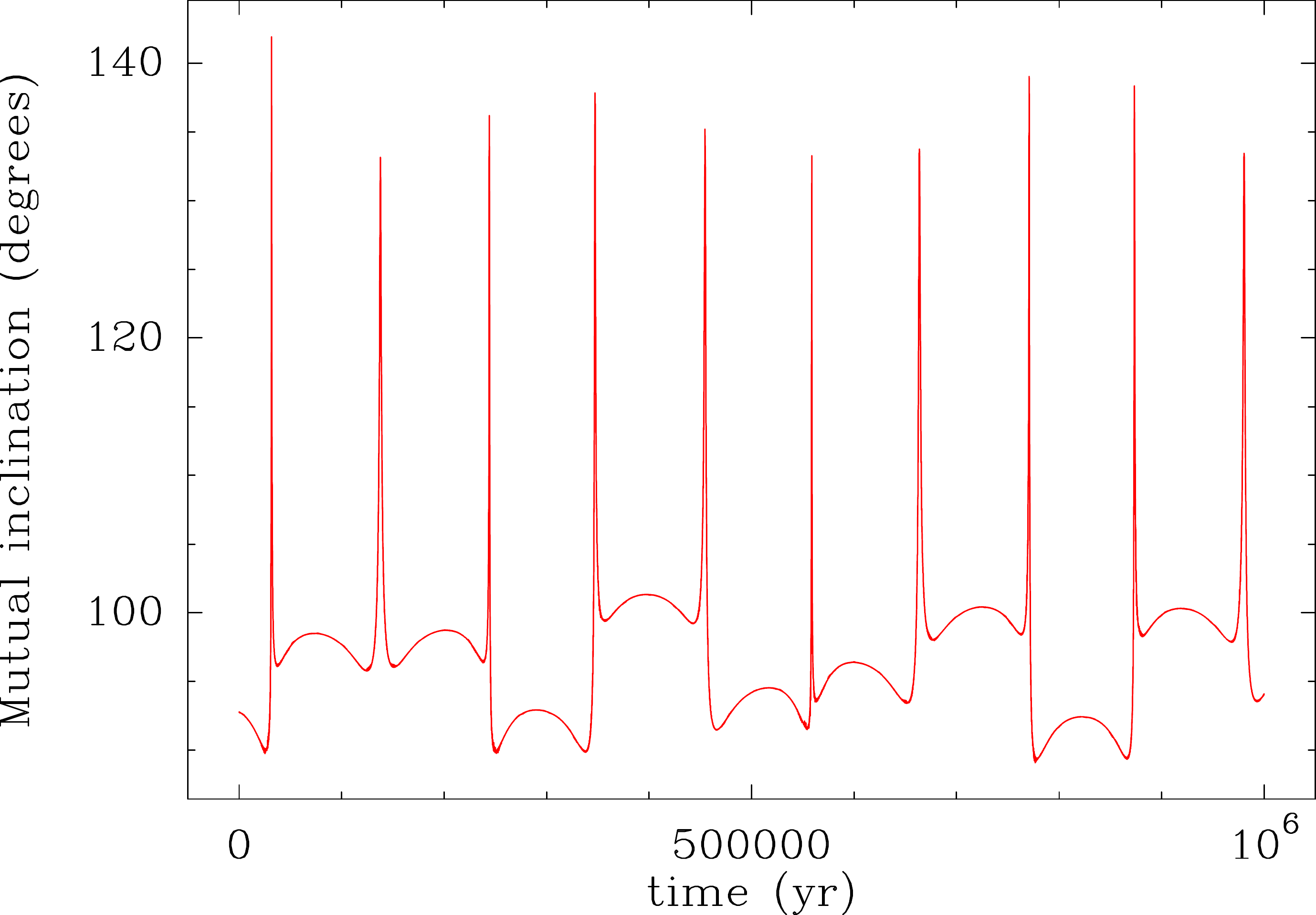} \hfil
\includegraphics[width=0.33\textwidth]{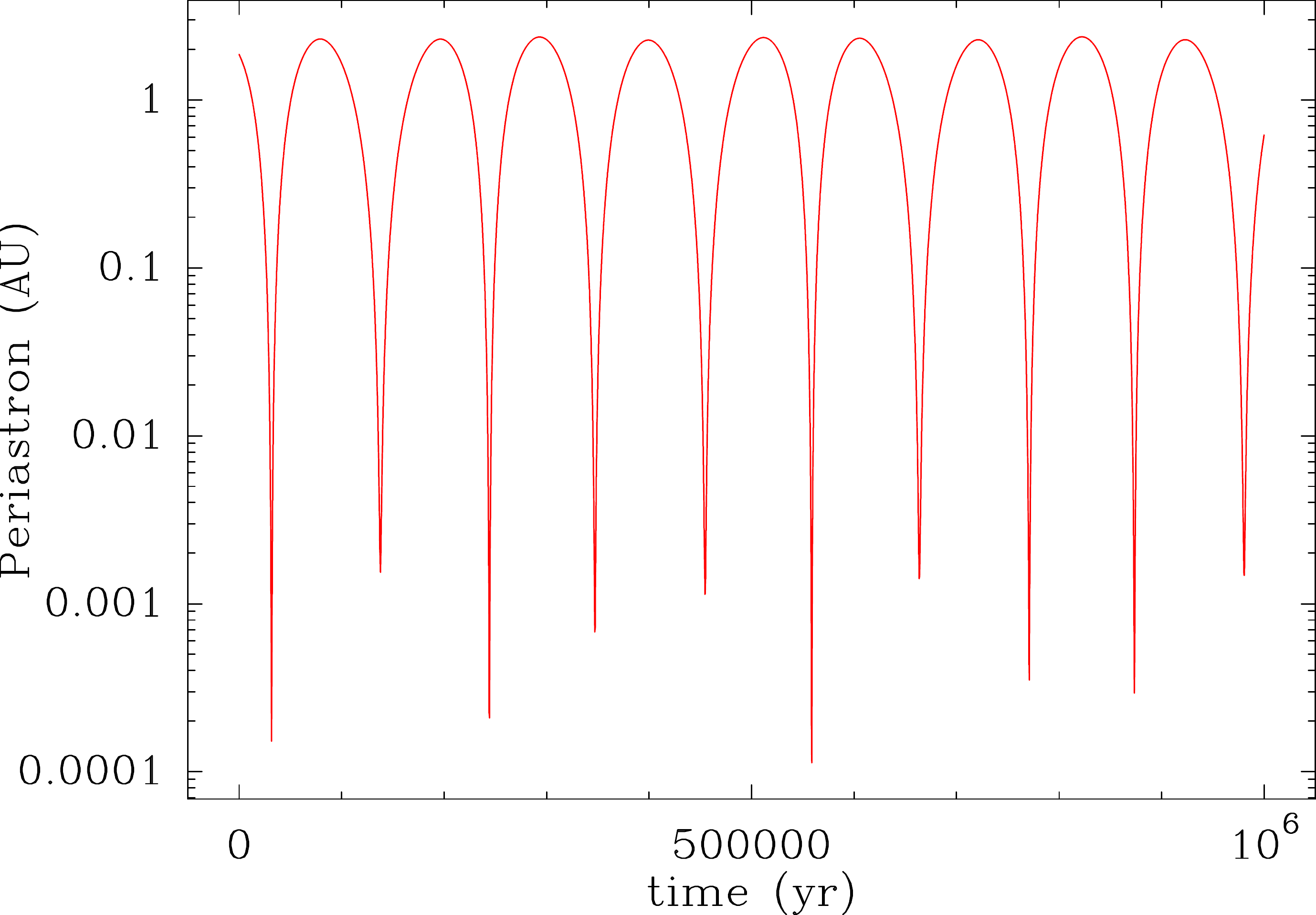}}
\caption[]{Orbital evolution of the PZ Tel$+$c system, as computed with pure 
Newtonian dynamics over 1 Myr, using HJS integrator. \textbf{Left :}
eccentricities of both orbits; red is for \pzc\ and green for \pz; 
\textbf{Middle :} Mutual inclination between both orbital planes; 
\textbf{Right} Periastron of \pzc, in logarithmic scale}
\label{pztelbc_hjs}
\end{figure*}
We thus decided to numerically investigate the dynamical stability of
this three body system, starting from the best $\chi^2$ solution for
\pz\ in Fig.~\ref{mcmcmap_pztelb_dk3}, and with the orbital solution
from Table~\ref{pztelc}. As the fitted orbit of \pz\ is barycentric,
we used our HJS symplectic code \citep{beu03} that naturally works in
Jacobi coordinates, which is the case here. The result is presented in
Fig.~\ref{pztelbc_hjs}, which shows the orbital evolution of the
system over $10^6$\,yr. The regular evolution pattern is an 
indication of
stability. In fact, the integration was carried out up to $10^7$\,yr,
which reveals the same behaviour as in the first $10^6\,$yr. Moreover,
the semi-major axes of both planets (not shown here) appeared to be
remarkably stable, which confirms the stability. Nonetheless, \pzc's
eccentricity exhibits large amplitude oscillations couple with
oscillation of the mutual inclination between both orbital
planes. This behaviour is characteristic for a strong Kozai
resonance. It must be noted however that in high eccentricity phases,
the mutual inclination does not drop down to 0 but up towards
$180\degr$ (retrograde orbits). This is thus an example of retrograde
Kozai resonance.

This picture is however very probably erroneous. In fact, given the
almost perfectly perpendicular initial configuration of the orbits,
the Kozai cycles drive \pzc\ up to very high eccentricity values. The
peak eccentricity in the cycles is actually close to $\sim 0.998$ !
Considering that \pzc's semi-major axis is nearly constant, its
periastron must drop down to very small values during peak
eccentricity phases. The right plot of Fig.~\ref{pztelbc_hjs} confirms
this fact. The minimum periastron value in the peaks ranges between
$10^4\,$AU and $10^{-3}\,$AU. With a mass of $1.25\,\msun$, PZ~Tel's
radius can be estimated to $9\times10^5\,$km, i.e,
$6\times10^{-3}\,$AU. \pzc\ is thus potentially subject to collision
with the central star. However, when the periastron decreases in high
eccentricity phases, \pzc\ is very probably affected by tides with the
central star that may prevent physical collisions. Tides were not
taken into account in the run described in Fig.~\ref{pztelbc_hjs}, so
that this picture does not hold.

\begin{figure*}
\makebox[\textwidth]{
\includegraphics[width=0.33\textwidth]{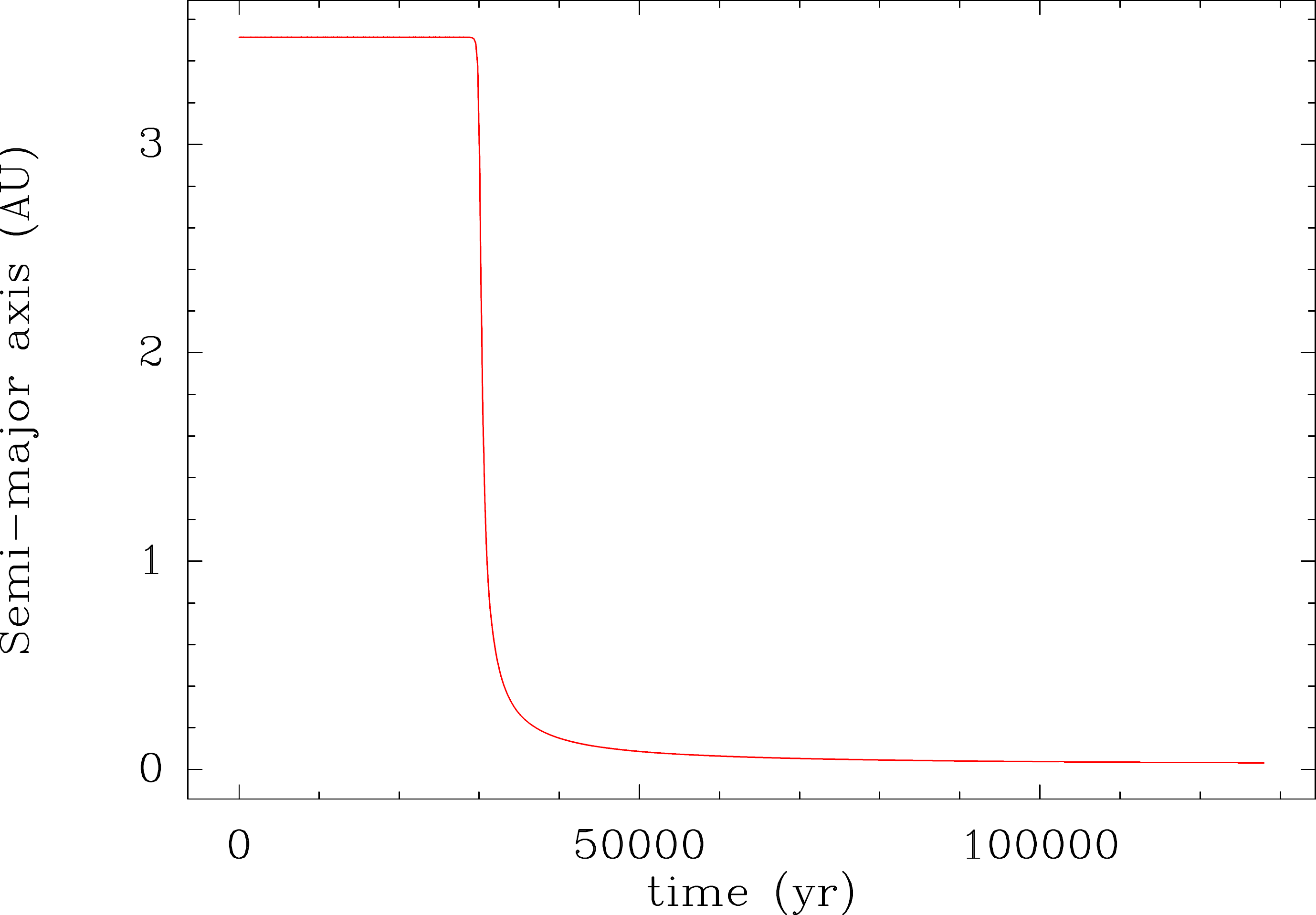} \hfil
\includegraphics[width=0.33\textwidth]{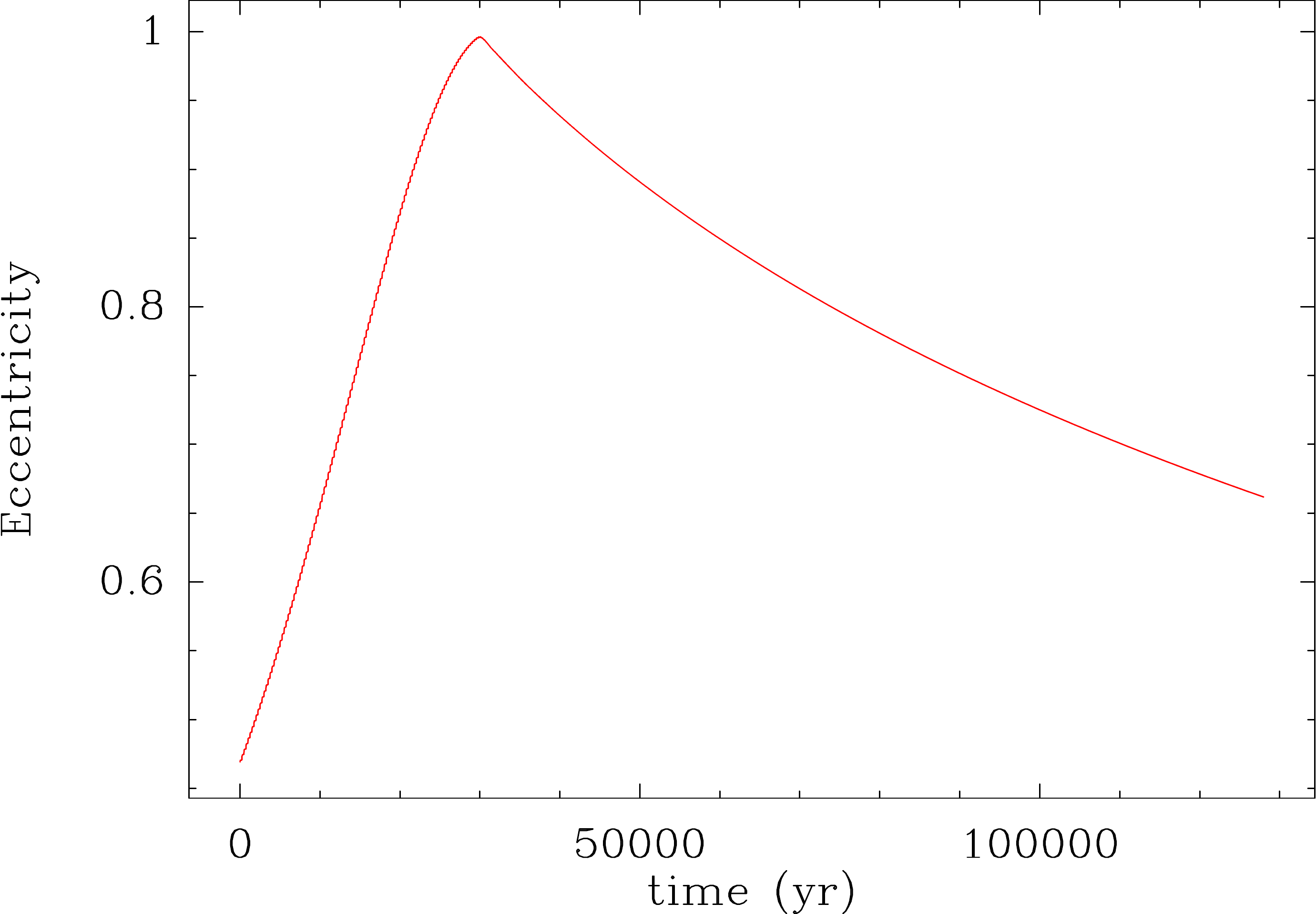} \hfil
\includegraphics[width=0.33\textwidth]{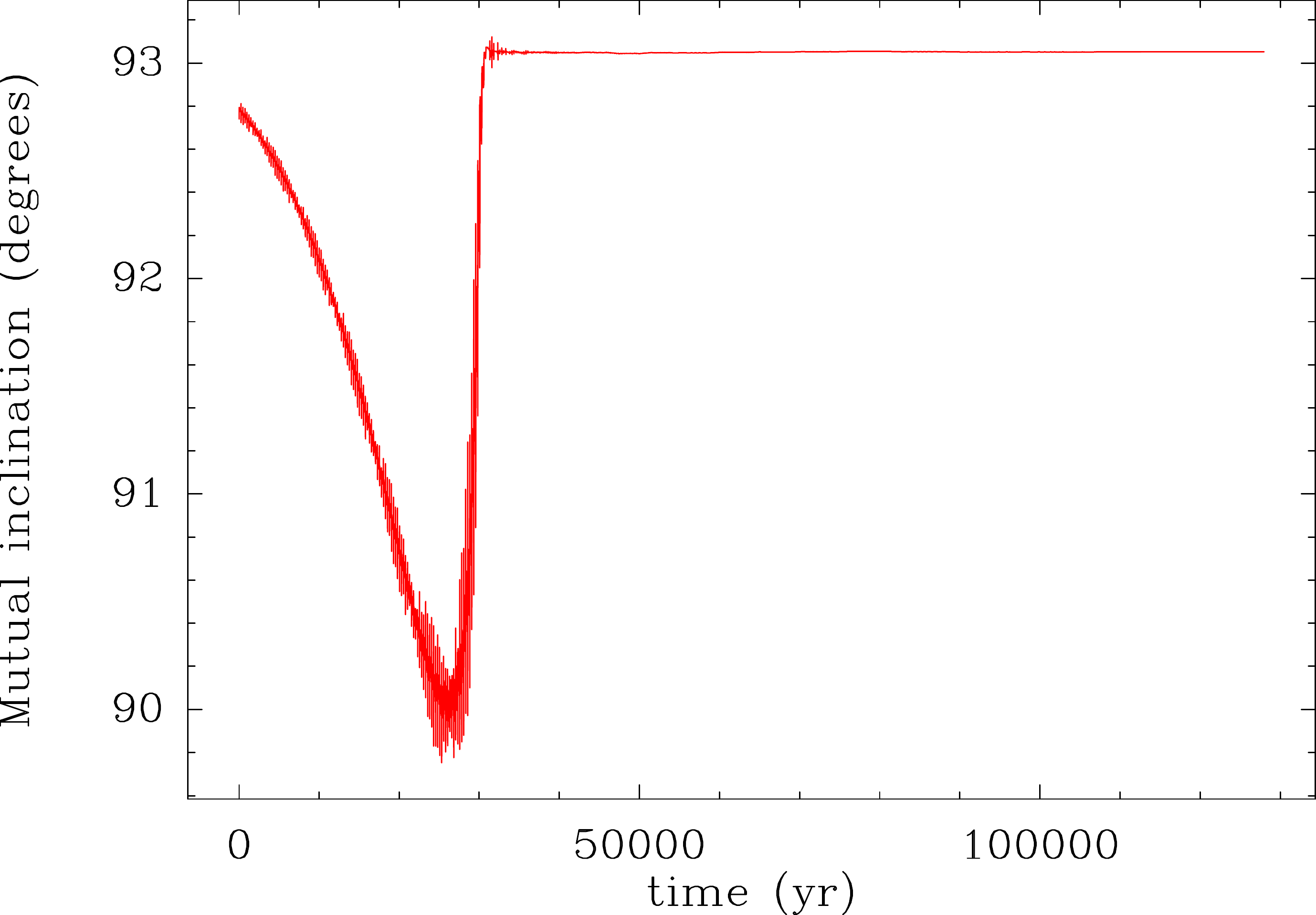}}
\caption[]{Orbital evolution of the \pzc\ system in a similar configuration
as in Fig.~\ref{pztelbc_hjs} with respect to \pz, but with tides taken
into account, assuming $Q=10^5$ quality factor. The computation was made
using the HJS integrator to which tides and post-Newtonian corrections
were added. \textbf{Left :} semi-major axis; \textbf{middle :}
eccentricity; \textbf{right :} Mutual inclination between both
orbital planes}
\label{pztelbc_tid}
\end{figure*}
We thus re-computed the secular evolution of the same three-body
system, but taking now tides between the central star and \pzc\ into
account. The computation was done using a special version of the HJS
integrator to which we have added tides and relativistic
post-Newtonian corrections. The details of this code are presented in
\citet{beu12}, with an application to the GJ~436 system. Tides mainly
depend of the assumed tidal dissipation parameter $Q_p$ for the
planet, a dimensionless parameter related to the rate of energy
dissipated through tides per orbital period \citep{bar09}. The smaller
$Q_p$, the more efficient the tidal dissipation is. $Q_p$ is very hard
to estimate, but typical values for giant planets range around $10^5$
within one order or magnitude \citep{zha08}.

Figure~\ref{pztelbc_tid} shows the result of such a simulation
assuming $Q_p=10^5$ for \pzc. The difference with the previous run is
striking. \pzc's semi-major axis appears to remain constant for
$~\sim3\times10^4,$yr and to suddenly drop after. During the first
phase (before $3\times10^4\,$yr), the eccentricity gradually increases
before decreasing, and the mutual inclination decreases before
stabilising. The explanation is the following: in the first phase, the
periastron remains too large to allow tides to be active. But the
Kozai mechanism causes an eccentricity increase until a point where
tides act at periastron. The subsequent energy dissipation causes then
a rapid decay of \pzc's orbit and a subsequent circularisation. This
scenario is actually similar to the one depicted in \citet{beu12} for
GJ~436 and by \citet{fab07}. In the latter cases, several Kozai-Lidov
oscillations together with tidal friction first occur before the inner
orbit starts to decay. Here this state is reached at the very first
eccentricity peak, thanks to very strong tides when the periastron
gets very close to the stellar surface.

The consequence is that tides prevent the deduced orbital
configuration between \pz\ and \pzc\ from being stable, as the
hypothetical \pzc\ inevitably migrates to a much closer orbit after
only $3\times10^4\,$yr. This result obviously depends on the assumed
$Q_p$ value. We thus tried other runs with increased $Q_p$ values
($10^6$ and above) to reduce the strength of tides. This appeared to
only delay the time of the orbital decay without changing the basic
scenario. In any case, \pzc\ ends up on a tight orbit ($<0.1\,$AU) on
a timescale in any case much lower than the age of the system.

Our conclusion is that the \pzc\ scenario depicted here to account for
the apparent very high eccentricity of \pz\ does not hold, as it would
require an orbital configuration that cannot be stable. We are thus
back to our conclusion that \pz's orbit is really very close to
parabolic state, as deduced from our initial MCMC analysis.
\section{Conclusion}
We have developed a new MCMC code based on the use of universal
Keplerian variables, dedicated to the orbital fit of imaged companions
with very high eccentricities or unbound orbits. This code was
successfully applied to the specific cases of \fomb\ and
\pz. Concerning \fomb, we confirm our orbital determination of
\citet{beu14}, but we show that the eccentricity distribution can
extend above $e=1$ in the unbound regime. This is in agreement with
the analysis of \citet{pea15} who show that for such companions imaged
over a very small orbital arc, the unknown radial velocity renders
unbound orbits possible. We think however that \fomb\ is very probably
a bound companion, although very eccentric. The case of \pz\ is more
complex. Our code reveals a very different eccentricity distribution
than for \fomb. \pz's eccentric distribution exhibits indeed a very
sharp peak around $e=1$.

According to \citet{pea14}, imaged companions can appear much more
eccentric than they are actually, thanks to the presence of hidden
inner companions that affect that astrometric data. \citet{pea14}
already showed that this model cannot account for \fomb's
eccentricity. Concerning \pz, we show that a hidden$\sim 12\,\mjup$
companion orbiting at $\sim 3.5\,$AU in a pole-on configuration
(contrary to the edge-on orbit of \pz) could mimic an almost unbound
orbit despite a real eccentricity around $\sim 0.7$. However, due to
the combination of tides and Kozai-Lidov mechanism, this configuration
is dynamically unstable. We are thus back to the conclusion that \pz\
is really a very high eccentricity, possible unbound companion.

The dynamical origin of \fomb\ and its configuration relative to the
dust disk orbiting Fomalhaut was recently investigated by
\citet{far15}. According to this model, the Fomalhaut system should
harbour another, more massive planet that controls the shape of the
disk. \fomb\ could have formerly resided in a mean-motion resonance
with that planet have been put on its present day eccentricity via a
gradual eccentricity increase and one or several close encounters. The
case of \pz\ is more complex. According to our orbital determination,
its eccentricity is in any case very close to 1 if not above. The
planet passed through a very close ($<0.1\,$AU) periastron in 2002,
consistent with its non-detection in 2003 NaCo images
\citep{mas05}. Highly eccentric orbits with very small periastron
are usually triggered by Kozai-Lidov mechanism or by mean-motion
resonance with a moderately eccentric outer companion \citep[such an
for instance in the so-called Falling Evaporating Body scenario in the
$\beta\:$Pictoris system][]{bm00}. But given the estimated mass of
\pz, this would require a perturbing companion in the low stellar mass
regime. Such a companion would have already been detected. Apart from
a very peculiar past encounter event, there is therefore no obvious
explanation for the unusual orbital configuration of \pz. Further
monitoring of this system will help better understanding it.
\begin{acknowledgements}
Most of the computations presented in this paper were performed
using the Froggy platform of the CIMENT infrastructure
(https://ciment.ujf-grenoble.fr), which is supported by the
Rh\^one-Alpes region (GRANT CPER07\_13 CIRA), the OSUG@2020 labex
(reference ANR10 LABX56) and the Equip@Meso project (reference
ANR-10-EQPX-29-01) of the programme Investissements d'Avenir
supervised by the Agence Nationale pour la Recherche.

The MCMC code used in this paper can be obtained freely upon request to 
\texttt{Herve.Beust@obs.ujf-grenoble.fr}.
\end{acknowledgements}
\end{document}